\documentclass[twocolumn]{aastex701}
\pdfoutput=1

\newcommand{\logg} {\log \textsl{\textrm{g}}}

\newcommand{\Te} {T_{\rm eff}}

\newcommand{\msun} {$M_\odot$}

\newcommand\gta{\lower 0.5ex\hbox{$\buildrel > \over \sim\ $}} %greater than about
\newcommand\lta{\lower 0.5ex\hbox{$\buildrel < \over \sim\ $}} %less than about
\newcommand{\nh} {{\rm H}/{\rm He}}
\newcommand{\halpha} {$\rm{H}{\alpha}$}

\newcommand{\hei} {He {\sc i}}

\newcommand{\sn}{{\rm S/N}}

\begin{document}

\title{A Theoretical Investigation of \hei\ Line Profiles for the
  Spectroscopic Analysis of DB White Dwarfs}

\author{Patrick Tremblay} 
\affiliation{D\'epartement de Physique, Universit\'e de Montr\'eal, C.P.~6128, Succ.~Centre-Ville, Montr\'eal, Qu\'ebec H3C 3J7, Canada}
\email[show]{patrick@astro.umontreal.ca}

\author{Pierre Bergeron} 
\affiliation{D\'epartement de Physique, Universit\'e de Montr\'eal, C.P.~6128, Succ.~Centre-Ville, Montr\'eal, Qu\'ebec H3C 3J7, Canada}
\email[show]{bergeron@astro.umontreal.ca}

\author{Alain Beauchamp} 
\affiliation{D\'epartement de Physique, Universit\'e de Montr\'eal, C.P.~6128, Succ.~Centre-Ville, Montr\'eal, Qu\'ebec H3C 3J7, Canada}
\email{alainbeauchamp1905@outlook.com}

\shortauthors{Tremblay et al.}
\shorttitle{DB White Dwarfs Revisited}

\begin{abstract}
We present a comprehensive investigation of \hei\ line profile
calculations used in the spectroscopic analyses of DB white
dwarfs. Our study includes an in-depth photometric and spectroscopic
analysis of all DB white dwarfs in the Data Release 17 of the Sloan
Digital Sky Survey, examining the effects of frequency sampling,
Doppler broadening, line dissolution, broadening by neutral particles,
and 3D hydrodynamical corrections on our results. More importantly, we
compare the outcomes obtained from the semi-analytical \hei\ Stark
profiles commonly used in DB white dwarf spectroscopic analyses with
our recent calculations of Stark-broadened profiles derived from
computer simulations.
\end{abstract}


\section{Introduction}

There are two primary methods for measuring the stellar parameters of
white dwarf stars. The first, known as the spectroscopic method
\citep{Bergeron1992,Finley1997,Liebert2005,Bergeron2011}, involves
comparing observed spectra with predictions from synthetic models to
determine the effective temperature ($\Te$) and surface gravity
($\logg$) of each object, as well as its atmospheric composition,
including any trace elements. The resulting $\Te$ and $\logg$ values
are then converted into stellar mass ($M$) using evolutionary
sequences for white dwarfs (e.g., \citealt{Fontaine2001}, \citealt{Bedard2020}), which
provide the necessary mass-radius relations. An alternative approach,
known as the photometric method
\citep{Bergeron1997,Bergeron2001,Bergeron2019,GentileFusillo2019},
relies on photometric observations (magnitudes), which are converted
into averaged fluxes using appropriate photometric calibrations
\citep{Holberg2006}. These fluxes are then compared with synthetic
flux predictions, averaged over each bandpass. In this method, the two
fitted parameters are $\Te$ and the solid angle $\pi(R/D)^2$, where
$R$ is the white dwarf's stellar radius, and $D$ is its distance from
Earth. The distance may be determined directly from trigonometric
parallax measurements, or alternatively, a specific value for the
white dwarf mass (or surface gravity) can be assumed. Once again, the
$\Te$ and $R$ values can be converted into mass using white dwarf
mass-radius relations. In certain cases, a hybrid spectroscopic and
photometric approach is applied, such as for DZ and DQ stars, where
the photometric method is used in conjunction with atmospheric
composition constraints derived from spectroscopic fits to optical
spectra \citep{Dufour2005,Dufour2007}.

Both methods have been successfully applied for the past 30 years or
so (see \citealt{Bedard2024} for a review), although the number of
objects available for analysis was initially quite limited. For
example, \citet{Bergeron1992} employed the spectroscopic method to
analyze a sample of only 129 DA stars in order to determine their mass
distribution. Likewise, \citet{Bergeron2001} used the photometric
method on a sample of 152 cool white dwarfs, utilizing optical $BVRI$
and infrared $JHK$ photometry as well as published trigonometric
parallax measurements. At the time, their relatively small sample
represented nearly all the parallax measurements available in the
literature. Today, however, thanks to large spectroscopic surveys like the Sloan
Digital Sky Survey (SDSS, \citealt{York2000}), extensive photometric
surveys like SDSS and the Panoramic Survey Telescope and Rapid
Response System (Pan-STARRS, \citealt{Chambers2016}), and
most importantly, the Gaia astrometric survey \citep{Gaia2018},
spectroscopic and photometric analyses of white dwarf stars have
advanced significantly, allowing thousands of objects to be analyzed
simultaneously
\citep{Bergeron2019,Kilic2020,McCleery2020,Caron2023,OBrien2024}.

These studies have, for the first time, enabled a detailed comparison
of the spectroscopic and photometric methods for determining the
stellar parameters of white dwarfs, allowing for assessments of their
validity, accuracy, and precision. Such in-depth comparisons were
carried out for DA and DB white dwarfs by \citet{Genest2019a} and
\citet{Tremblay2019}. Although the measurements generally agree,
significant discrepancies were noted, potentially due to issues with
photometric calibrations or limitations in the model atmospheres,
particularly in line-broadening theory and the treatment of convective
energy transport.

In this paper, we focus on DB white dwarfs, whose spectra are
dominated by neutral helium lines, and may also contain weak hydrogen
lines (primarily \halpha) in the case of DBA
stars. \citet{Genest2019b} presented what is arguably the most
extensive and comprehensive photometric and spectroscopic study of the
DB white dwarf population to date, analyzing $ugriz$ photometry and
medium-resolution spectroscopy from the SDSS, combined with
trigonometric parallax measurements from the Gaia mission. The authors
conducted a detailed comparison of the $\Te$, $\logg$, and mass
measurements derived from both photometric and spectroscopic
techniques to evaluate the precision and accuracy of each method. The
most relevant findings for our discussion are shown in their Figures 7
and 10, which display the $\logg$ and mass distributions,
respectively, as functions of $\Te$ for both methods. The photometric
distributions exhibit the expected mean values of
$\langle\logg\rangle\sim8$ and $\langle M\rangle\sim0.6$ \msun\ across
the entire temperature range, while the spectroscopic distributions
show considerable deviations from these means at most
temperatures. Several explanations were proposed for these
discrepancies. For instance, above $\Te\sim27,000$ K, where
spectroscopic masses appear too low, residual flux calibration issues
with the SDSS spectra were suggested as a possible cause. This issue
had also been reported by \citet{Genest2014}, who observed a similar
phenomenon in the spectroscopic $\logg$ distribution of DA white
dwarfs from the SDSS. At lower temperatures ($22,000~{\rm
  K}>\Te>16,000$ K), the spectroscopic $\logg$ values for DB stars
appear overestimated relative to the mean. This issue could be
partially mitigated by applying the 3D hydrodynamical corrections of
\citet{Cukanovaite2018}, who developed 3D model atmospheres for pure
helium-atmosphere white dwarfs using the CO$^5$BOLD
radiation-hydrodynamics code. The corrected $\logg$ values, shown in
the bottom panel of Figure 7 in \citet{Genest2019b}, show improvement
down to around 19,000 K; however, at lower temperatures, the
spectroscopic $\logg$ values appear significantly underestimated,
suggesting that the 3D corrections may have been overestimated. It was
noted, however, that these 3D corrections were initially available
only for pure helium atmospheres, and that calculations including
hydrogen could potentially resolve this issue.

More recently, \citet{Cukanovaite2021} revisited this issue by
calculating a comprehensive set of 3D corrections that included model
atmospheres accounting for the presence of hydrogen. Their findings
showed that the 3D $\logg$ corrections were most substantial for
$\Te\lesssim18,000$ K, reaching up to $-0.2$ dex at $\logg =
8.0$. However, since $\logg$ determinations in this temperature range
could also be significantly influenced by the treatment of broadening
by neutral particles, as well as non-ideal effects from helium
perturbation by neutral atoms, it remains challenging at this stage to
draw firm conclusions regarding the validity of these 3D
corrections. Further improvements in understanding the microphysics of
helium are needed to reduce these uncertainties.

In this spirit, we decided to undertake an investigation by
reevaluating the physics of line broadening theory. As a first step,
we began by refining the Stark broadening calculations for neutral
helium lines by \citet[][hereafter B97]{beauchamp97}, which are the
most widely used tables in the spectroscopic analyses of DB white
dwarfs. These calculations rely on the standard Stark broadening
theory with the inclusion of the one-electron theory in the far wings
(see below). \citet{TremblayP2020} introduced new calculations of
Stark-broadened profiles for neutral helium lines using computer
simulations that unify the treatment of ions and electrons. These
simulations also incorporate corrections for ion dynamics, transitions
of electron contributions from the core to the wings of the profile,
numerical integration of the time evolution operator for helium
perturbed by a fluctuating electric field, the Debye correction for
charge-motion correlations, local density variations, and particle
reinjection. \citet[][hereafter Paper I]{Tremblay2026} have applied a
similar, but more advanced, framework to generate a comprehensive set
of Stark profile calculations for most neutral helium lines relevant
to the analysis of DB white dwarf optical spectra in the 3800 to 5200
\AA\ wavelength range.

In this paper, we present a comprehensive approach to understanding
the effects of line broadening on the spectroscopic analysis of DB
white dwarfs. In Section \ref{stark}, we summarize the key physical
components of line broadening theory for neutral helium, with
particular emphasis on the Stark broadening of \hei\ line
profiles. Section \ref{analysis} describes the theoretical framework
and observational data used to test the impact of various elements
within the line broadening theory, which are then individually
examined in Section \ref{results}. Our conclusions follow in
Section \ref{conclusion}.

\section{Calculation of \hei\ Line Profiles}\label{stark}

The spectroscopic technique used for DB stars typically relies on
optical spectra, where \halpha, if present, helps to constrain
the hydrogen abundance, while the blue portion of the spectrum is used to
determine $\Te$ and $\logg$ (see, e.g., \citealt{Bergeron2011}). In
this section, we briefly discuss some of the physical components
involved in calculating model spectra for determining the atmospheric
parameters of DB white dwarfs.

The spectra of DB white dwarfs are primarily dominated by neutral
helium lines, with occasional traces of hydrogen, mostly as
\halpha. The main broadening mechanism for \hei\ lines in these stars
is Stark broadening, caused by charged perturbers in the surrounding
plasma. The most widely used calculations for Stark-broadened profiles
were done by B97 over 25 years ago, and these have remained the
standard in the field.  These profiles are computed using the standard
Stark broadening theory as implemented by \citet{BCS69} and
\citet{BC70}, with the inclusion of the one-electron theory of
\citet{Baranger62}; see also \citealt{TremblayP2020} and Paper I for a
review of the theory and a discussion on these various approximations.
Stark profiles for 21 \hei\ lines are provided as tables covering a
range of electronic densities ($N_e=10^{14}$ to $6\times10^{17}$
cm$^3$), temperatures (10,000 K, 20,000 K, and 40,000 K), and grids of
40 to 70 frequency points. Convolution with Doppler broadening is also
included.

Another crucial aspect of these calculations is the inclusion of line
dissolution. In brief, rather than extending the integral over the
ionic field to infinity, it is carried out only up to the critical
value at which the highest Stark levels for a given $n$ merge with
other levels (see equations 4.1 and 5.5 of B97
following the method originally proposed by \citet{Seaton1990}. In
this approach, the critical electric field is calculated using the
theory introduced in the occupation probability formalism by
\citet{HM1988}. Note that a similar method was used by
\citet{Tremblay2009} in their Stark broadening calculations for
hydrogen lines.

As discussed above, \citet[][and Paper I]{TremblayP2020} presented new calculations
of Stark-broadened profiles for \hei\ lines using computer simulations
that eliminated several approximations found in the earlier work of
B97, with the most significant improvement being the
inclusion of ion dynamics. However, line dissolution has not yet been
incorporated into these calculations. We evaluate these effects
separately in Section \ref{results}.

Below $\Te\sim16,000$~K, \hei\ line broadening by neutral particles
begins to dominate over Stark broadening. The implementation of van
der Waals and resonance broadening in our model calculations is
described at length in \citet{BeauchampPhD} and summarized in Section
3.2 of \citet{Genest2019a}. Unfortunately, broadening by neutral particles
introduces significant uncertainty in estimating white dwarf masses
using the spectroscopic technique. This is illustrated in Figure 11 of
\citet{Genest2019a}, which compares the spectroscopic mass
distributions of DB white dwarfs in the SDSS across effective
temperatures, using the van der Waals broadening theory of
\citet{Unsold1955} and \citet[][see Section \ref{vdW} below for
  details]{Deridder1976}. It is challenging to determine which theory
is more suitable until other issues, including the impact of 3D
hydrodynamical corrections on spectroscopic $\logg$ determinations,
are fully addressed. Complicating matters further, at such low
temperatures, \hei\ lines eventually disappear, rendering the
spectroscopic technique unreliable. At this point, the photometric
technique should be used instead.

\section{Spectroscopic and Photometric Analysis}\label{analysis}

The spectroscopic and photometric data for our analysis are taken from
the Data Release 17 (DR17) of SDSS. We selected all DB spectra with a
signal-to-noise ratio above $\sn=20$, resulting in a total of 738
objects. All magnetic white dwarfs were excluded from our
analysis. For the photometric analysis, we combined $ugriz$ photometry
with Gaia DR3 parallax measurements. Our model atmospheres are
detailed in \citet{Bergeron2011}, with updates to
broadening by neutral particles discussed in \citet{Genest2019a}. The model spectra are
calculated using various Stark-broadened profiles for \hei\ lines,
which are described in the following section. Note that in our
comparative analysis, the temperature and pressure structures remain
identical across all cases, as the specifics of the line profile
calculations have no significant effect on the atmospheric structure
--- something we verified retrospectively.

Our aim in this paper is to explore various approximations within the
\hei\ broadening theory. To achieve this, we will compare the
spectroscopic results, $\Te$ and $\logg$, obtained under different
approximations to better understand the systematic effects introduced
by each physical component. Additionally, we will examine the global
properties of the DB sample by comparing photometric and spectroscopic
mass distributions as a function of effective temperature.

\begin{figure}
% x1 y1 x2 y2
\includegraphics[clip=true,trim=1cm 4cm 0.5cm 1.5cm,width=\columnwidth]{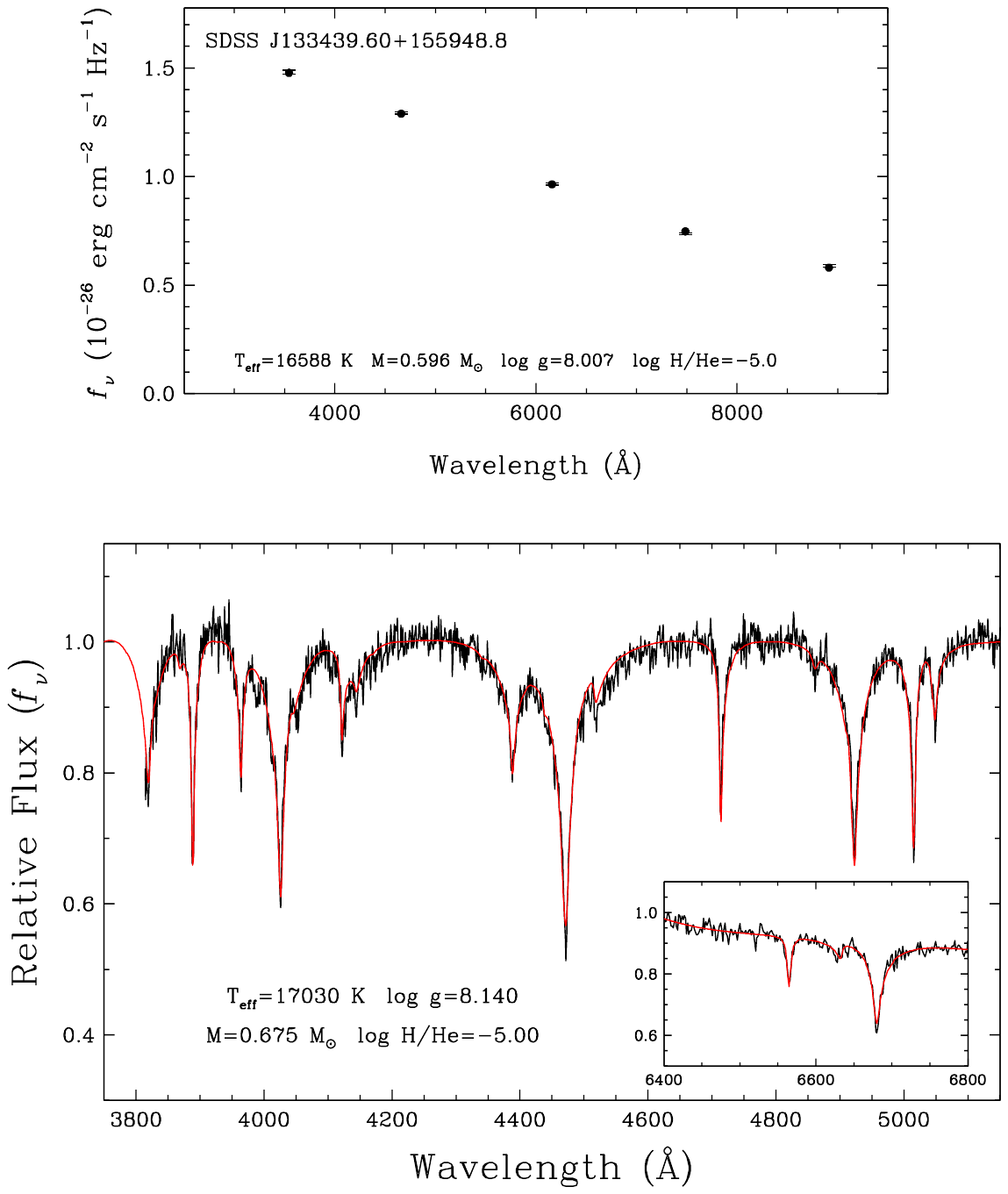}
\caption{Best photometric and spectroscopic fits to a typical DB white
  dwarf in the DR17 SDSS sample. In the top panel, the observed
  $ugriz$ photometry is shown as error bars (1$\sigma$ uncertainty),
  while the model atmosphere fit is shown as solid dots, with the
  parameters given in the panel. The H/He abundance ratio in this fit
  is constrained from the spectroscopic fit below. The lower panel
  shows our spectroscopic fit, with the observed and model spectra
  displayed in black and red, respectively, both normalized to a
  continuum set to unity.  The spectroscopic fit to the H$\alpha$ line
  profile, shown in the insert, is used to measure, or constrain, the
  hydrogen abundance of the overall solution.
\label{samplefit}}
\end{figure}

Our determinations of the photometric and spectroscopic parameters are
illustrated in Figure \ref{samplefit}. In the top panel, the $ugriz$
photometry, converted into absolute fluxes using the AB calibration,
is compared to model spectra predictions, averaged over each filter
bandpass. Interstellar reddening is accounted for using STILISM values
\citep{Lallement2014,Capitanio2017}. The fitted parameters here are
the effective temperature, $\Te$, and the solid angle, $\pi(R/D)^2$,
where $R$ is the star's radius and $D$ is its distance from Earth,
obtained from the Gaia parallax measurement. The radius is then
converted into stellar mass using the cooling sequences from
\citet{Bedard2020} with C/O-core (50/50) compositions, $q({\rm
  He})\equiv M_{\rm He}/M_{\star}=10^{-2}$, and $q({\rm H})=10^{-10}$,
appropriate for non-DA stars. We first fit the spectral energy
distribution assuming a pure helium composition. The resulting
photometric temperature is then used as an initial estimate to fit the
SDSS spectrum, thereby avoiding the issue of choosing between the
typically ambiguous cool and hot solutions that arise when using only
the spectroscopic technique --- one on either side of the peak
strength of the neutral helium lines, which occurs near 25,000 K for
DB white dwarfs (see Section 4.2 of \citealt{Bergeron2011} for further
discussion). The measured hydrogen-to-helium abundance ratio (H/He by
number) is then used to obtain updated estimates of the photometric
parameters.

The spectroscopic technique applied to DB/DBA white dwarfs follows the
method described in detail by \cite{Bergeron2011} and relies on
normalized spectra, as illustrated in the bottom panel of Figure
\ref{samplefit}. Briefly, the normalization procedure for DB spectra
utilizes our synthetic spectra and a high-order polynomial (up to
$\lambda^5$) to produce a smooth fitting function for determining the
continuum location. Unlike the approach in Bergeron et al., here we
use the photometric temperature to determine on which side of the
maximum line strength the solution lies. Once the observed spectrum is
normalized, we obtain an initial estimate of $\Te$ and $\logg$ using
the blue portion of the spectrum ($3800-5200$ \AA). Keeping these
atmospheric parameters fixed, we then determine (or constrain) the
hydrogen abundance by fitting the region near H$\alpha$ ($6400-6800$
\AA). This process is repeated until the ${\rm H}/{\rm He}$ value
converges. When H$\alpha$ is undetectable, only upper limits on the
hydrogen abundance can be obtained. The $\logg$ value is then
converted into stellar mass using the same cooling sequences as
before. As noted above, the spectroscopically determined H/He value is
also used for the photometric fit displayed in the top panel of Figure
\ref{samplefit}.

With these determinations, we can now make various comparisons between
different \hei\ profiles using the stellar parameters derived from
both the spectroscopic and photometric methods.

\section{Results}\label{results}

\subsection{Reappraisal of the Semi-Analytical \hei\ Line Profiles}\label{reapp}

Since the publication of our Stark profiles in
B97, we have continuously maintained and improved our
code, although no new calculations have been officially
released. Specifically, the code has been rewritten from Fortran to
C++, and minor errors have been identified and corrected. Calculation
precision has been enhanced by redefining a small number of constants,
originally specified as single-precision, to
double-precision. Additionally, the convergence criteria for numerical
integrals (over ionic microfields and for convolution with a Doppler
profile) have been tightened, and the interpolation procedure for
Hooper tables \citep{Hooper68}, which describe the microfield
distribution as a function of thermodynamic conditions, has been
improved. Two new options have been added to the code. The effect of
line dissolution can now be included (or excluded) in the calculation
of profiles by controlling the upper limit of the integral over the
ionic microfield. Moreover, the electron contribution to broadening
can either be described by impact broadening for the full profile or
switch to the one-electron regime in the wings. The originally
published profiles included line dissolution and combined both
electronic regimes. To ensure direct comparison between the
semi-analytical and simulation methods presented in this article, the
mixing of the upper transition levels due to interactions with the
electric field of the perturbers (ions and electrons) is the same for
both sets of profiles, and has been restricted to levels with the same
principal quantum number $n$ as the upper level of the allowed
transition.

\begin{figure}
% x1 y1 x2 y2
\includegraphics[clip=true,trim=1.5cm 2cm 1.5cm 1.4cm,width=\columnwidth]{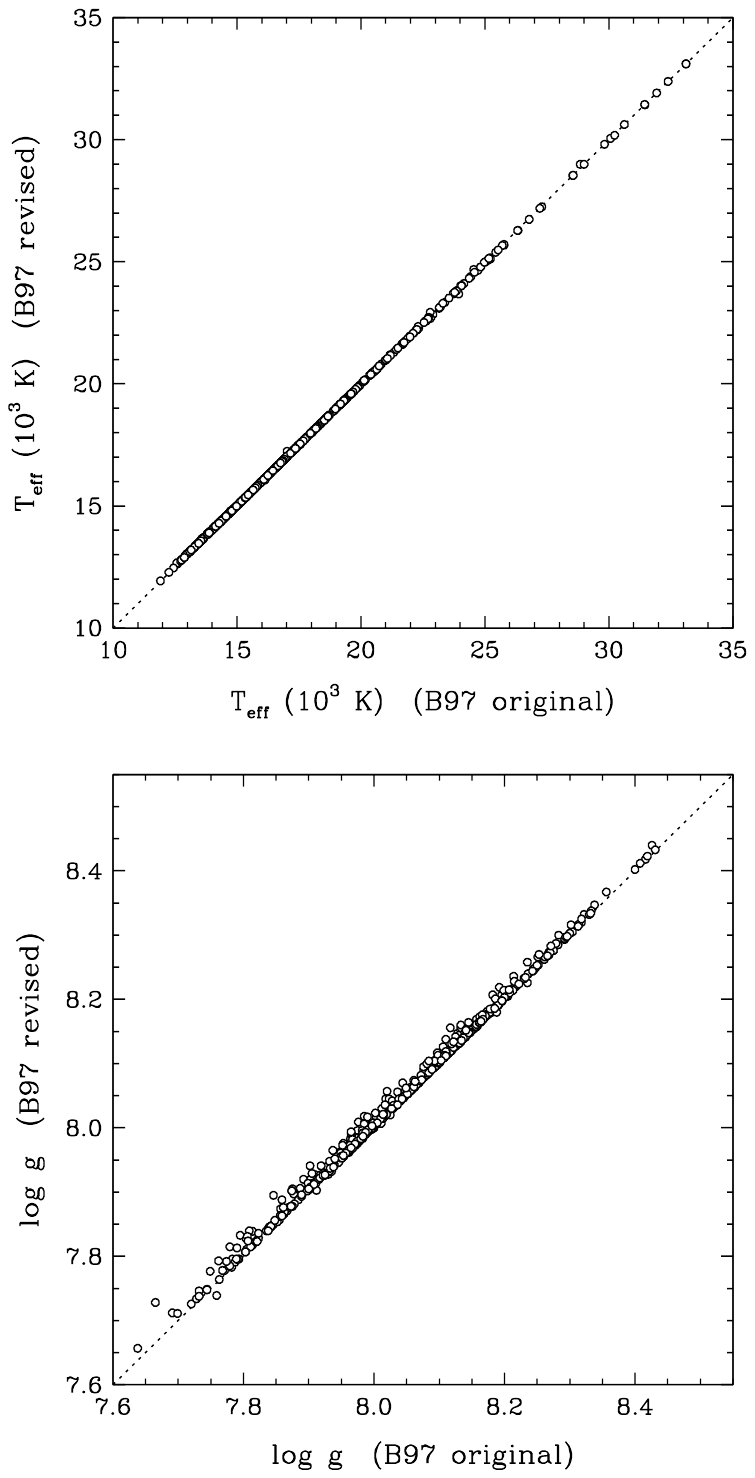}
\caption{Comparison of effective temperatures and surface gravities
  for the DB white dwarfs in the DR17 SDSS sample obtained from
  spectroscopic fits using the Stark profile calculations from
  B97 and our revised calculations, as
  described in the text. The dotted lines indicate the 1:1
  correspondence.
\label{comptg0}}
\end{figure}

We first present a comparison of the results obtained between both
sets of calculations. Figure \ref{comptg0} compares the $\Te$ and
$\logg$ values for the DB white dwarfs in the DR17 SDSS sample
obtained from spectroscopic fits using the Stark profile calculations
from B97 alongside our updated calculations described above. Both
sets of calculations consider line dissolution, Doppler broadening,
resonance and van der Waals broadening, following a modified version of the
theory by \cite{Deridder1976} as described in Section
\ref{vdW}. Additionally, both profile sets rely on tables with
identical frequency sampling. Note that in Figure \ref{comptg0} (and
subsequent figures), we have restricted the range of $\Te$ and $\logg$
values to the field of interest, although objects exist outside this
range, particularly at higher effective temperatures.

\begin{figure}
% x1 y1 x2 y2
\includegraphics[clip=true,trim=1.2cm 9cm 1.2cm 2cm,width=\columnwidth]{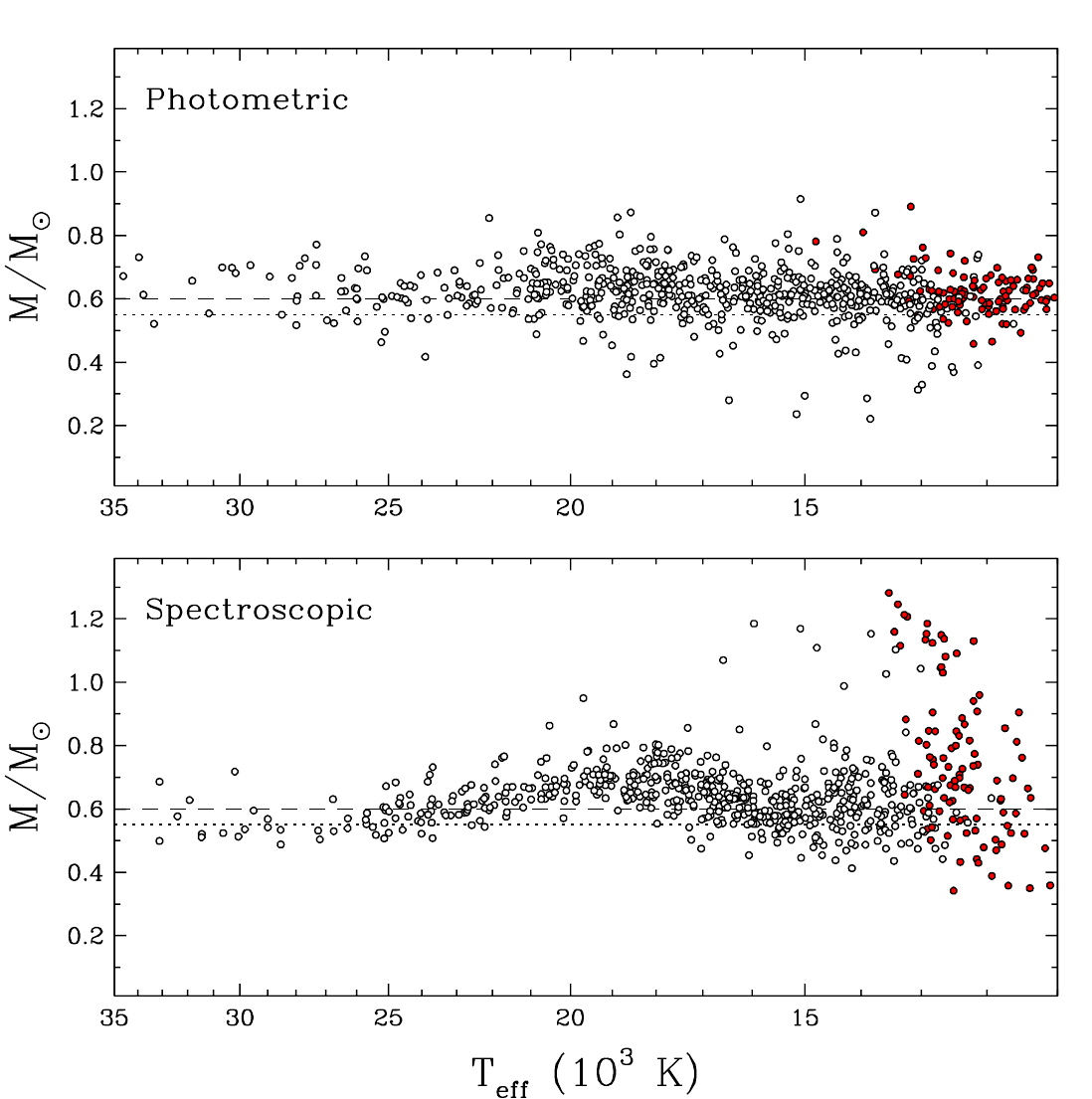}
  \caption{Photometric (upper panel) and spectroscopic (bottom panel)
    mass distributions as a function of $\Te$ for the DB white dwarfs
    in the DR17 SDSS sample with $\sn>20$. The spectroscopic
    parameters have been obtained using model spectra that include our
    revised Stark broadening profiles (see text and Figure
    \ref{comptg0}) based on the approach described in B97. The objects
    with an equivalent width ${\rm EW}< 5$ \AA\ for
    \hei\ $\lambda$4471 are shown by red symbols.  In both panels, the
    dashed and dotted lines correspond to a constant mass of 0.6 and
    0.55 \msun, respectively.
    \label{correltm1}}
\end{figure}

The comparison shown in Figure \ref{comptg0} indicates that our
revised calculations are nearly identical to those of B97, with the
updated $\Te$ values being, on average, only 2.5 K lower and the
updated $\logg$ values just 0.007 higher. Figure
\ref{correltm1} further presents the spectroscopic mass distribution
as a function of $\Te$ for the same sample, comparing it with the
photometric mass distribution. As a reference, we show the canonical
mean mass of 0.6 \msun\ as a dashed line in each panel and a constant
mass of 0.55 \msun\ as a dotted line. The latter corresponds to the
empirical mass threshold below which common-envelope evolution must be
invoked to explain the presence of low-mass white dwarfs, based on the
photometric mass distribution of over 4000 white dwarfs analyzed by
\citet[][see their Figures 17, 23, and 24]{Kilic2025}. Consequently,
white dwarfs evolving through a single-star channel should lie above
this mass limit, as is the case for most DB white dwarfs observed in
the photometric mass distribution (upper panel of Figure
\ref{correltm1}).

The results displayed in Figure \ref{correltm1} are entirely
consistent with those presented in Figures 8 and 11 of
\citet{Genest2019a} and, more importantly, Figure 7 of \citet[][in
  this case showing the $\logg$ distribution]{Genest2019b}. We briefly
summarize the conclusions of these studies here, and direct the reader
to Section 5.1 of \citet{Genest2019b} for a more detailed discussion.
While the photometric mass distribution remains well-centered on the
canonical mean mass of approximately 0.6 \msun\ for white dwarfs
across the full $\Te$ range shown in Figure \ref{correltm1}, the
spectroscopic mass distribution deviates significantly from the 0.6
\msun\ value in several temperature ranges. For $\Te \gtrsim
27,000$~K, the low spectroscopic masses are thought to be artifacts
due to calibration issues with the SDSS spectra, a problem also noted
by \citet{Tremblay2011} and \citet{Genest2014} in the context of DA
white dwarfs. In the range $22,000~{\rm K} > \Te > 16,000$~K, the
spectroscopic masses appear overestimated relative to the photometric
values; this corresponds to a temperature range where 3D
hydrodynamical corrections are expected to play an important role, as
suggested by \citet{Cukanovaite2018,Cukanovaite2021}. Note that these
3D corrections have not been incorporated into our results and will be
addressed below. Finally, for $\Te \lesssim 14,000$~K, the
spectroscopic distribution shows a notable increase in both mass and
dispersion, especially when compared to the photometric mass
distribution. This high-mass or high-$\logg$ issue is typically
attributed to limitations in the broadening theory by neutral particles,
which we discuss further below.

We reexamined this high-mass issue using our analysis results and a
detailed inspection of our spectroscopic fits. It became evident that
the high-mass white dwarfs in Figure \ref{correltm1} correspond to
objects with the weakest helium lines. Similar to DA stars, below a
certain $\Te$ threshold --- also dependent on $\logg$ --- the
absorption lines weaken to the point where the spectroscopic technique
becomes unreliable. To quantify this effect for DB stars, we measured
the equivalent width (EW) around \hei\ $\lambda$4471 (from 4220 to
4625 \AA) for all objects in our sample. The DB stars with ${\rm EW} <
5$ \AA\ are highlighted in red in Figure \ref{correltm1}. Below this
threshold, \hei\ $\lambda$4471 becomes one of the few lines detectable
in the blue portion of the spectrum (as used in Figure
\ref{samplefit}). Excluding these objects from our analysis reveals in
Figure \ref{correltm1} that the high-mass issue almost completely
disappears. This conclusion, of course, likely depends on the
restricted ${\rm S/N} > 20$ range of our spectroscopic sample.

From this point forward, the B97 profiles will refer to the revised
calculations described in this section.

\subsection{Frequency Sampling and Doppler Broadening}\label{Doppler}

As explained in the Introduction, \hei\ Stark profiles are typically
provided in the form of tables across a range of electron densities,
temperatures, and a grid of frequency (or wavelength) points. Our
original tables contained 40 to 70 frequency points, depending on each
\hei\ line. In the course of our spectroscopic analysis, we discovered
that the precise number of frequency points in these tables was
particularly important, especially for forbidden transitions and very
narrow absorption lines such as \hei\ $\lambda$3889, which is notably
$\logg$-sensitive in cool DB white dwarfs. More importantly, we also
identified significant issues with the convolution of our profiles
with Doppler broadening, which we now discuss in detail.

\begin{figure}
% x1 y1 x2 y2
\includegraphics[clip=true,trim=1.8cm 8cm 1cm 8.5cm,width=\columnwidth]{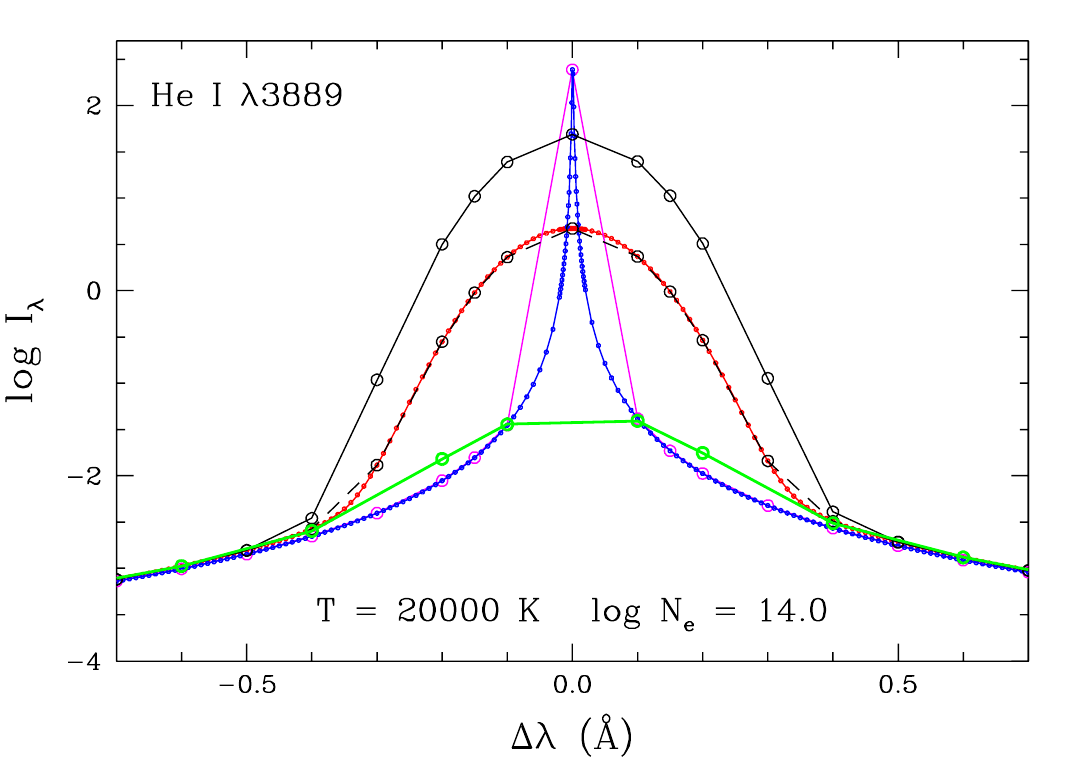}
\caption{Comparison of various Stark intensity profiles for He {\sc i}
  $\lambda$3889 at $T=20,000$~K and $N_e=10^{14}$ cm$^{-3}$. The blue line
  corresponds to the unconvolved profile sampled on an extremely fine wavelength
  grid marked by small blue dots. The red line (with small dots) is the
  same profile convolved with a Doppler profile, while the black
  dashed line corresponds to the resampling of this convolved profile on our
  2X wavelength grid (see text) marked by open circles. The
  magenta line corresponds to the unconvolved profile sampled on the
  2X wavelength grid marked by open circles, while the black solid
  line corresponds to the direct convolution of this profile with a
  Doppler profile (using the same wavelength grid). Finally, the green
  line shows the convolved line profile using the original 1X
  wavelength grid from B97 marked by open circles.
  \label{DopplerFigure}}
\end{figure}

Tables of Stark profiles are commonly provided both with and without
convolution with a Doppler profile (see, e.g., the case of hydrogen
lines by \citealt{vcs73}), which accounts for the atom’s velocity
along the line of sight \citep{mihalas78}. The tables of
B97 were only available to the scientific community
with Doppler broadening included. In our recent experiments using revised
calculations with the original (1X) frequency sampling from the tables
of B97 --- displayed in Figure \ref{comptg0} --- and
double (2X) the frequency sampling, we found that our Doppler profile
convolution was problematic, particularly for extremely narrow lines,
such as the $\logg$-sensitive \hei\ $\lambda$3889. The problem, along
with the solution, is illustrated in Figure \ref{DopplerFigure}, where
we show various intensity profiles for \hei\ $\lambda$3889 at
$T=20{,}000$~K and $N_e=10^{14}$ cm$^{-3}$.

The blue line in Figure \ref{DopplerFigure} shows the unconvolved
profile sampled on an extremely fine wavelength grid, indicated by
small blue dots (note that the wavelength range in this figure is
approximately 1 \AA\ wide). The convolution of this profile with a
Doppler profile is shown as the red line, with the same wavelength
sampling marked by red dots. As explained above and in Paper I, we now
favor twice the frequency sampling (2X) used in B97 to better
represent the core of some allowed and forbidden components of
\hei, especially at low densities. The resampling of the convolved
profile on this coarser grid is shown as a black dashed line, with
open circles indicating the wavelength sampling. The agreement with
the convolved profile on the fine wavelength grid (red line) is
excellent, demonstrating that the overall procedure is sufficiently
accurate, and this is the approach we now adopt.

The issue with the earlier calculations of B97 is illustrated
by the magenta profile in Figure \ref{DopplerFigure}, which shows the
unconvolved Stark profile sampled on the coarse 2X wavelength grid,
represented by open magenta circles. At this particular low electron
density, this undersampled profile is a poor representation of the
true profile shown in blue. As a direct consequence of this
undersampled profile, the resulting convolution with a Doppler profile
--- shown by the black solid line and black open circles --- yields a
higher intensity near the line core (and consequently a deeper
absorption line) when compared to our more accurate calculations
(dashed black line). The situation is even worse with the convolved
profiles using the original (1X) sampling of B97,
shown by the green line and open green circles; here, a crucial point
is even missing at $\Delta\lambda=0$, which further exacerbates the
problem.

\begin{figure}
% x1 y1 x2 y2
\includegraphics[clip=true,trim=1.5cm 2cm 1.5cm 1.4cm,width=\columnwidth]{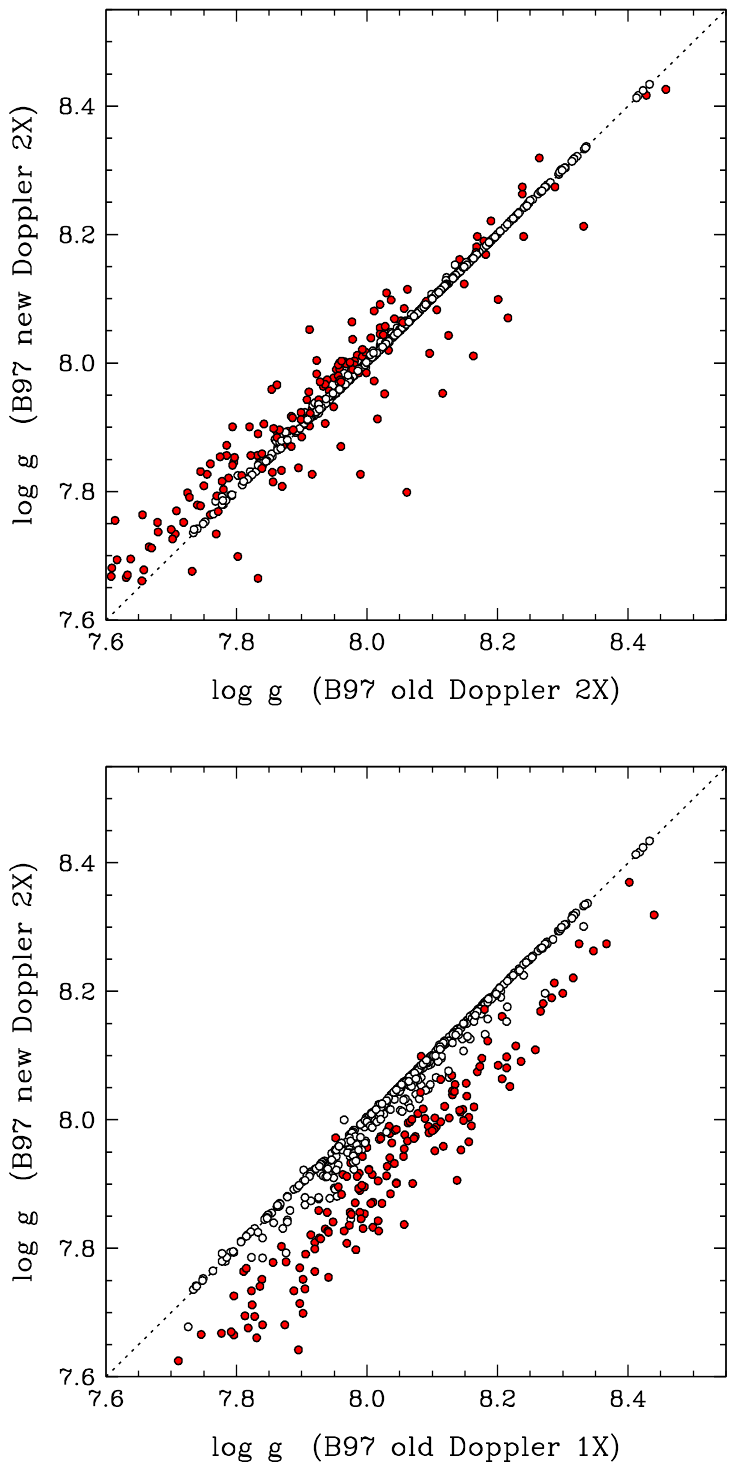}
\caption{Comparison of surface gravities for the DB white dwarfs in
  the DR17 SDSS sample obtained from spectroscopic fits using our revised
  version of our semi-analytical Stark profile calculations where
  Doppler broadening is included accurately (new Doppler 2X; see text)
  and from our previous version (old Doppler) with 1X (bottom panel)
  and 2X (top panel) the original frequency sampling of B97.
  Objects with $\Te<15,000$~K are shown in red, while the dotted line
  indicates the 1:1 correspondence.
\label{comptg1}}
\end{figure}

We now provide a detailed comparison of the atmospheric parameters of
DB stars in the DR17 SDSS sample using the various calculations
described in Figure \ref{DopplerFigure}. In Figure \ref{comptg1}, we
compare the $\logg$ values derived from synthetic spectra using the
B97 version of our calculations with corrected Doppler broadening
(labeled new Doppler 2X), as described above, and the earlier B97
version with inaccurate Doppler broadening (labeled old Doppler) ---
described in Section \ref{reapp} and used in Figure \ref{comptg0} ---
with the original (1X, bottom panel) and double (2X, top panel) the
frequency sampling used in the tables of B97; the effects on $\Te$ are
negligible and are therefore not shown. The comparisons indicate that
$\logg$ values differ most significantly for cool DB white dwarfs with
$\Te<15,000$~K, shown in red in both panels, a direct consequence of
our corrected treatment of Doppler broadening. The results with double
(2X) the frequency sampling (top panel) show a significant scatter in
this range of temperature, but with only a small systematic shift in
$\logg$. However, the effects become more pronounced when we compare
the results with the original (1X) frequency sampling (bottom
panel). In this case, our corrected treatment of Doppler broadening
predicts $\logg$ values that are, on average, approximately 0.1 dex
lower for cool DB white dwarfs.

\begin{figure}
% x1 y1 x2 y2
\includegraphics[clip=true,trim=1.cm 5.5cm 0.cm 3.7cm,width=\columnwidth]{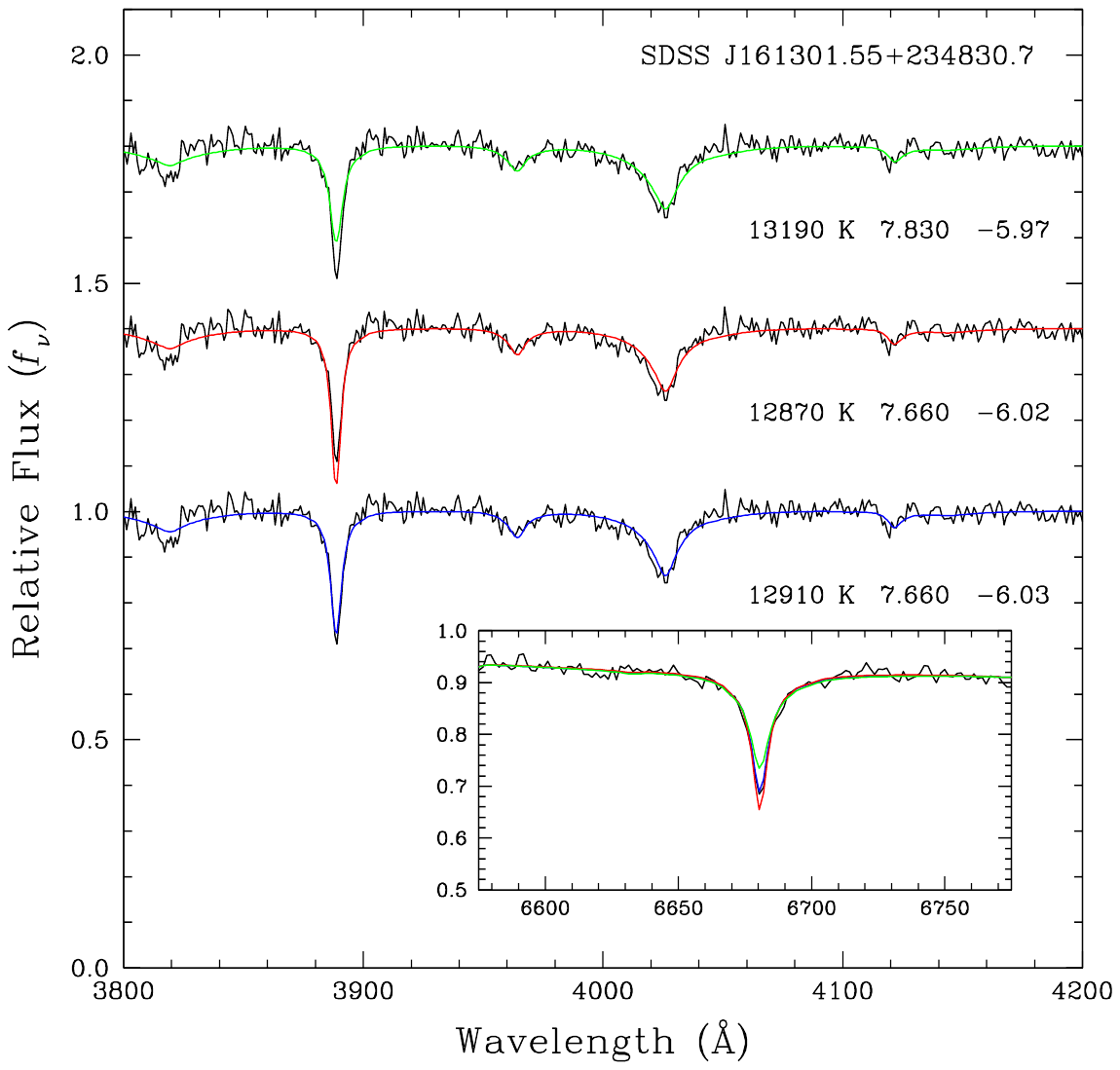}
\caption{Best spectroscopic fits to the DB white dwarf SDSS J161301.55+234830.7
  with model spectra calculated with the B97
  version of the Stark profiles with corrected Doppler broadening
  (bottom, blue) and the B97 version with the old (and inaccurate)
  Doppler broadening version with 2X (middle, red) and 1X (top, green)
  the frequency sampling of the original tables from B97. The values
  of $\Te$, $\logg$, and $\log\nh$ of each fit are given in the
  figure. The insert shows the corresponding region around
  \hei\ $\lambda$6678.
\label{sampling}}
\end{figure}

We can gain more insight by comparing our best spectroscopic fits to a
specific cool DB white dwarf using the various calculations described
in Figure \ref{DopplerFigure}. We show such a fit in Figure
\ref{sampling} using the calculations with the proper
inclusion of Doppler broadening and frequency sampling (2X, blue line), and the improper inclusion
of Doppler broadening with 1X (green line) and 2X (red line) the original
frequency sampling of B97. The most significant
differences between these fits occur for \hei\ $\lambda$3889 and
$\lambda$6678. As seen for \hei\ $\lambda$3889, our corrected
calculations accurately predict the depth of the absorption profile,
whereas the 1X and 2X profiles are predicted to be too shallow and too
deep, respectively, consistent with the intensity profiles shown in
Figure \ref{DopplerFigure}. A similar conclusion applies to
\hei\ $\lambda$6678. For this particular object, the $\logg$ value
obtained from our corrected profiles is the same as that obtained with
the 2X grid. However, the surface gravity measured using the 1X grid
is 0.17 dex higher, in line with the systematic shift observed in the
bottom panel of Figure \ref{comptg1}. The reason for this difference
is that synthetic spectra based on the 1X frequency sampling (green
profile in Figure \ref{DopplerFigure}) are simply unable to reproduce
the observed deep line core, even by lowering the $\logg$ value; the
solution is thus driven by other lines in the spectrum.

To summarize this subsection, we adopt in what follows improved
semi-analytical \hei\ Stark profiles calculated using twice (2X) the
frequency sampling of the original grid of B97, including additional
frequencies tightly packed around the line core (as shown by the blue
dots in Figure \ref{DopplerFigure}). This grid is then used to
calculate the convolution with a Doppler profile (red
dots in Figure \ref{DopplerFigure}), and for the final grid, resampled
on the 2X frequency grid (the black dashed line and open
circles in Figure \ref{DopplerFigure}). This has been done for all
lines, and not just for \hei\ $\lambda$3889.

\subsection{Line Profile Normalization}\label{norm}

The semi-analytical Stark profiles from B97 include quasi-static
broadening by ions together with electron broadening. The latter is
treated using the impact approximation near the line core and the
one-electron approximation in the far wings, where strong collisions
play a more significant role. An interpolation scheme is applied to
ensure a smooth transition between the one-electron and impact
regimes. As discussed extensively in Paper~I, while the Stark profiles
derived solely from the impact theory are, by definition, properly
normalized, this is not the case when the one-electron approximation
is used. This lack of normalization was identified in Paper~I as one
of the main sources of discrepancy between our semi-analytical
profiles and those obtained from our computer simulations, in
particular at high density. This finding eventually led us to improve
our semi-analytical profiles by implementing a detailed normalization
procedure (see Section~3.3 of Paper~I).

\begin{figure}
% x1 y1 x2 y2
\includegraphics[clip=true,trim=1.5cm 2cm 1.5cm 1.4cm,width=\columnwidth]{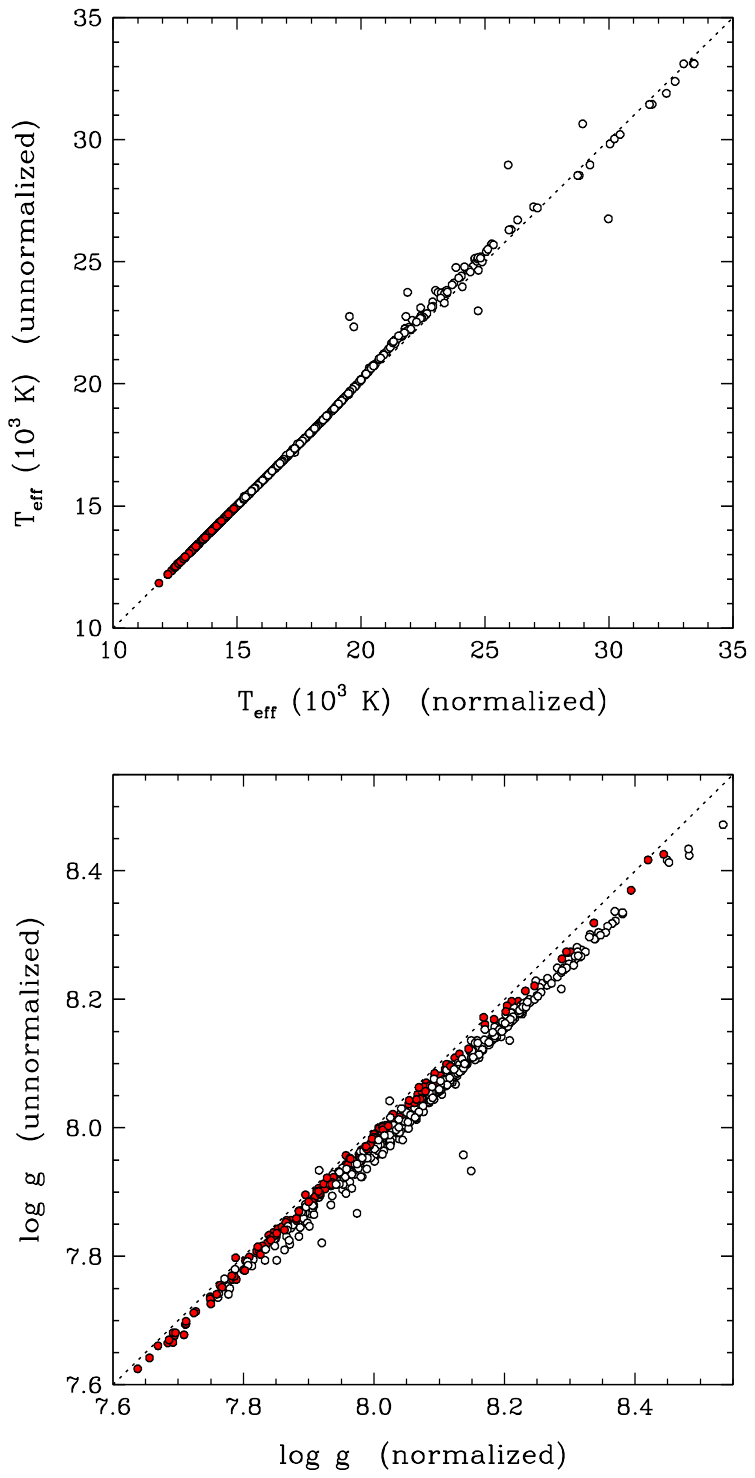}
\caption{Comparison of effective temperatures and surface gravities
for the DB white dwarfs in the DR17 SDSS sample obtained from
spectroscopic fits using the revised version of our semi-analytical
Stark profile calculations with and without normalization (see
text). Objects with $\Te<15,000$~K are shown in red, while the
dotted line indicates the 1:1 correspondence.
\label{comptg6}}
\end{figure}

We illustrate the impact of this normalization procedure in
Figure~\ref{comptg6}, which compares the effective temperatures and
surface gravities for the DB white dwarfs in the DR17 SDSS sample
obtained using spectra calculated from the unnormalized grid of
semi-analytical He~{\sc i} Stark profiles (as described above and used
previously) with those derived from the same profiles after proper
normalization. While the effects on the effective temperatures are
negligible, the $\logg$ values obtained from the normalized profiles
are, on average, about 0.03~dex lower, with the largest differences
observed for white dwarfs hotter than 15,000~K.

This normalization procedure represents the final improvement
implemented in the semi-analytical approach originally developed by
B97. We designate these improved Stark profiles as B25, which will
serve as our reference framework for the comparisons presented in the
remainder of this paper.

\subsection{Line Dissolution}\label{sec:linediss}

As briefly discussed in the Introduction, another important aspect of
line broadening calculations is the inclusion of line dissolution,
originally introduced by \citet{Seaton1990}. This concept was first
applied in the context of \hei\ Stark profiles by
B97. In summary, the most general form of the profile,
using the quasi-static approximation for ions and the impact
approximation for electrons, is the following average over the ion
microfield distribution of microfield-dependent electron-broadened
profiles \citep{Griem62,BCS69,BC70},

\begin{equation}
\label{eq:QS}
    I(\omega)  = \int_0^\infty dF \ W_a(F) \ I_e(\omega,F)\ , 
\end{equation}

\noindent
where $I_e(\omega,F)$ is the electron-broadened profile at frequency
$\omega$, $W_a(F)$ is the microfield distribution parameterized by $a$
--- the ratio of the mean distance between perturbers to the Debye
radius ---, and $F$ is the electric field. To account for the effects
of line dissolution, the upper boundary in this integral is replaced
by $F^{\rm crit}$, the critical field value at which the highest Stark
levels with a given $n$ merge with other levels. This critical field
value is obtained for the excited levels of the \hei\ atom using the
method of \citet{Seaton1990}, based on the theory of \citet{HM1988} in
the context of their occupation probability formalism (see B97 for
further details).  This treatment does not account for the additionnal
dissolution induced by interactions with neutral
particles. Incorporating this supplementary contribution would require
substantial modifications to the semi‑analytical theory, as this
effect cannot be captured through a simple truncation of the ionic
electric field. To our knowledge, no studies addressing this issue are
currently available. As mentioned previously, a similar approach was
used by \citet{Tremblay2009} in their Stark broadening calculations
for hydrogen lines (see their Section 3).  Note, however, that our
grid of model spectra are calculated using the occupation formalism of
\citet{HM1988} for both the hydrogen and helium populations and
corresponding bound–bound, bound–free, and pseudo-continuum opacities,
as implemented by \citet{Dappen1987}, and that in that context, the
occupation probabilities include the contributions of both charged and
neutral particles (see also Section IV of \citealt{Bergeron1991}).

\begin{figure}
% x1 y1 x2 y2
\includegraphics[clip=true,trim=1.8cm 1.2cm 2.cm 1.5cm,width=\columnwidth]{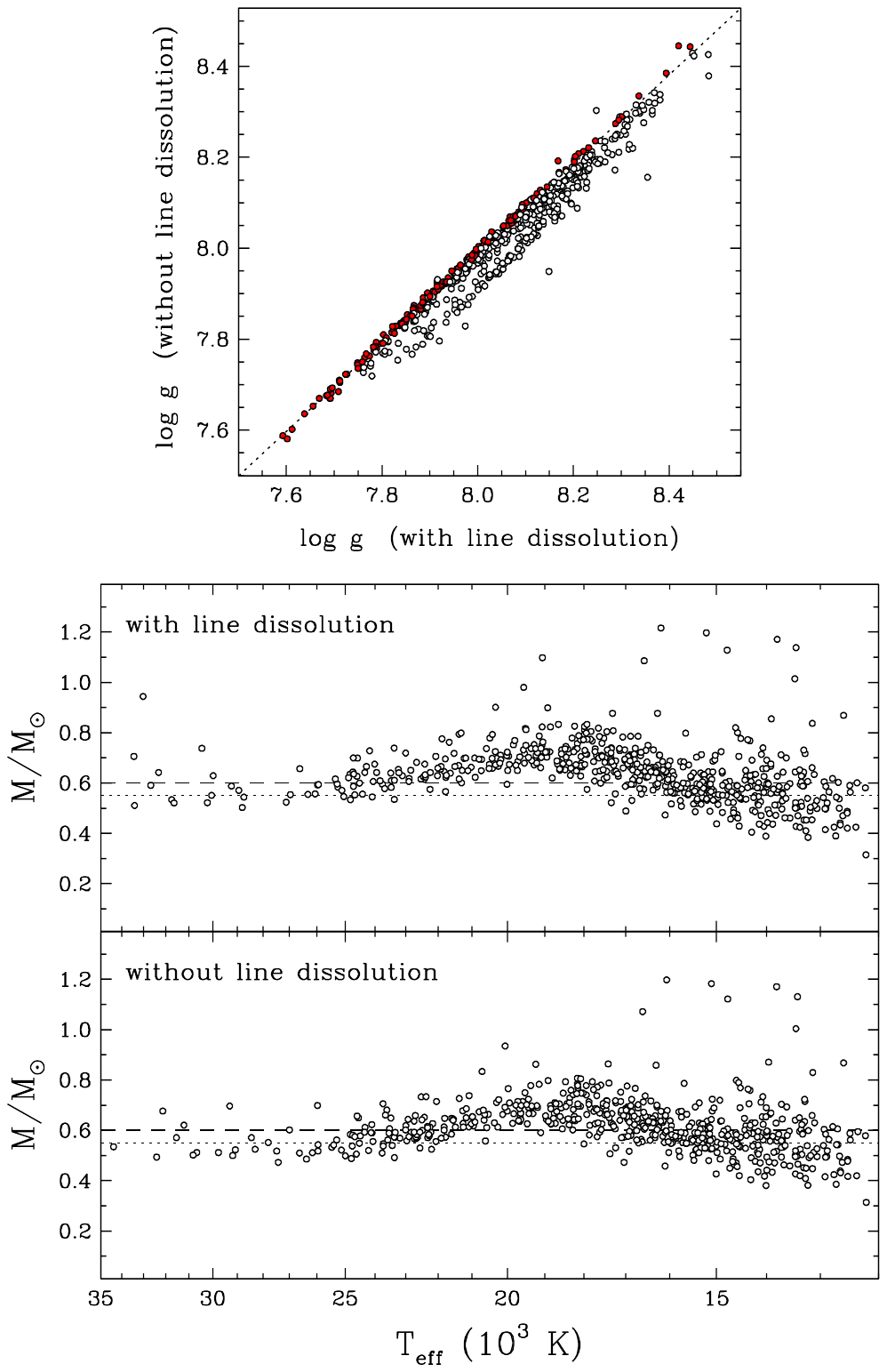}
\caption{Top panel: Comparison of surface gravities for the DB white
  dwarfs in the DR17 SDSS sample obtained from spectroscopic fits
  using our updated Stark profiles (B25) with and without line
  dissolution taken into account in the calculations; objects with
  $\Te<15,000$~K are shown in red, while the dotted line indicates the
  1:1 correspondence. Bottom panels: Spectroscopic mass distributions as
  a function of $\Te$ for the same sample using \hei\ profiles with and without
  line dissolution; the dashed and dotted lines correspond to a
  constant mass of 0.6 and 0.55 \msun, respectively.
\label{comptg3}}
\end{figure}

To study the effects of line dissolution on the \hei\ Stark profiles,
we calculated a table of line profiles similar to B25 but by excluding
this particular process. The spectroscopic $\logg$ values obtained
with and without line dissolution are compared in the top panel of
Figure \ref{comptg3}, while the bottom panels show the corresponding
mass distributions as a function of $\Te$; the spectroscopic $\Te$
values are not significantly affected and are therefore not compared
here. Our results indicate that neglecting line dissolution reduces
the $\logg$ values by only about 0.05 dex (or $\sim$0.03 \msun\ in
mass), and this effect is limited to hot ($\Te>15,000$~K) DB white
dwarfs. Although the effect of line dissolution is not negligible, it
remains small compared to other sources of uncertainty.

\subsection{\hei\ Line Profiles from Computer Simulations}

In the semi-analytical approach of B97 for calculating
the Stark profiles of neutral helium, electrons were treated as
dynamic and ions as static. With advancements in computational power,
other research groups have significantly enhanced the treatment of
Stark broadening by conducting computer simulations that detail the
dynamics and interactions of the perturbers (ions and electrons) near
the emitter --- see in particular \citet{Gig85}, \citet{Gig87}, and
\citet{Gig96} in the case of the hydrogen atom. Stark profiles for the
helium atom, however, were generated for only two lines,
\hei\ $\lambda$4471 and $\lambda$4922 \citep{Gig09,Lara12},
limiting their applicability to the computation of synthetic spectra
for white dwarf stars.

In \citet{TremblayP2020}, we introduced our own simulation environment
for calculating \hei\ Stark profiles, incorporating several key
aspects from previous work to better represent the dynamical
environment of the helium atom. These aspects include the unification
of ion and electron treatments, corrections for ion dynamics, the
proper treatment of electron broadening from the core to the wing,
still assuming the semi-classical approximation, the numerical
integration of the quantum operator governing the time evolution of
helium perturbed by a fluctuating electric field, the Debye correction
for the correlation of charged perturbers’ motions, as well as local
density variations and particle reinjection. Intermediate steps, such
as constructing the simulation space and the quasi-static model,
allowed us to validate the simulation space and ensure the system's
statistical consistency.

In that exploratory work, we generated grids of Stark profiles for the
two most important optical lines of the helium atom, namely
\hei\ $\lambda$4471 and $\lambda$4922, for temperatures between 10,000
K and 40,000 K and electron densities ranging from
$1\times10^{14}\,\mathrm{cm}^{-3}$ to
$6\times10^{17}\,\mathrm{cm}^{-3}$. More recently, in Paper I, we
improved these simulations and produced an extensive grid of Stark
profiles for 12 neutral helium lines covering the same range of
temperatures and electron densities, and a 13th line
(\hei\ $\lambda$4713) with the same temperatures but with electron
densities from $3\times10^{15}\,\mathrm{cm}^{-3}$ to
$6\times10^{17}\,\mathrm{cm}^{-3}$ due to its narrow nature, as
explained in detail in Paper I. The key improvement in these
calculations was the application of the power spectrum equation for
line profile generation, rather than the autocorrelation function used
in the earlier calculations of \citet{TremblayP2020}. This helped to
reduce the noise level in our calculations, especially in the far
wings, as confirmed mathematically by \cite{Rosato2020} and
demonstrated by \cite{Cho2022} for hydrogen lines, thus enhancing the
time-quality ratio of the method.  This advancement enabled, for the
first time, more realistic calculations of synthetic spectra for white
dwarf stars. However, the process of line dissolution, as described in
Section \ref{sec:linediss}, was not included in these line profile
calculations.

\begin{figure}
% x1 y1 x2 y2
\includegraphics[clip=true,trim=1.5cm 2cm 1.5cm 1.4cm,width=\columnwidth]{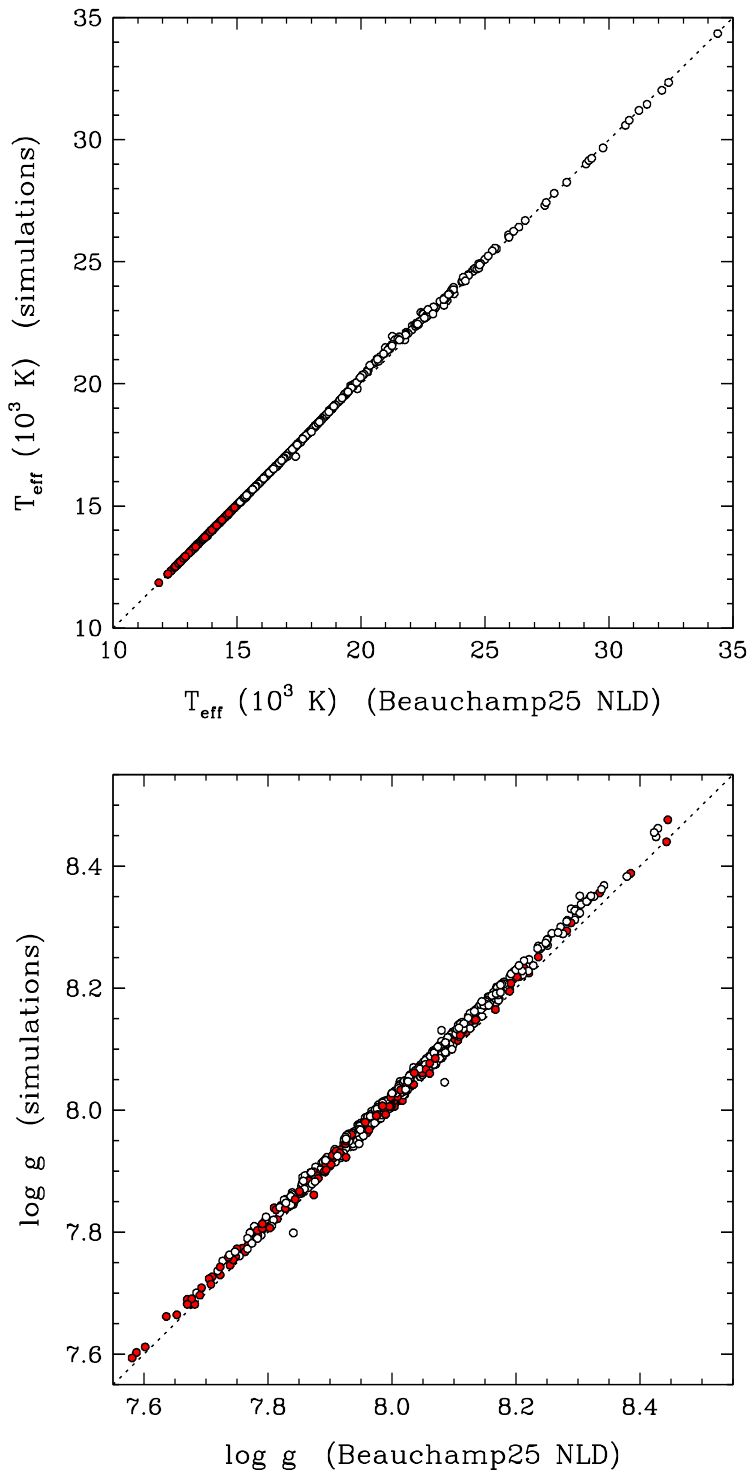}
\caption{Comparison of effective temperatures and surface gravities
  for the DB white dwarfs in the DR17 SDSS sample obtained from
  spectroscopic fits using the B25 Stark profiles with no line
  dissolution (NLD) and the computer simulations from Paper~I. Objects
  with $\Te<15,000$~K are shown in red, while the dotted line
  indicates the 1:1 correspondence.
\label{comptg4}}
\end{figure}

We compare in Figure \ref{comptg4} the effective temperatures and
surface gravities of the DB white dwarfs in the DR17 SDSS sample,
using \hei\ profiles based on our new computer simulations and the B25
Stark profiles. The Deridder \& van Rensbergen formalism for van der
Waals broadening (see Section \ref{vdW}) is applied in both sets of
calculations. Given our simulation-based profiles do not account for
line dissolution, we also exclude this process in the B25 calculations
(labeled as NLD in the following figures, for “no line
dissolution”). Our results show that the effective temperatures are
nearly identical, with an average difference of less than 130~K. The
$\logg$ values, however, display a systematic offset, with the Stark
profiles derived from computer simulations yielding surface gravities
about 0.03~dex higher than those obtained from the semi-analytical
profiles, regardless of $\Te$.

\begin{figure}
% x1 y1 x2 y2
\includegraphics[clip=true,trim=1.cm 6.0cm 0.cm 3.0cm,width=\columnwidth]{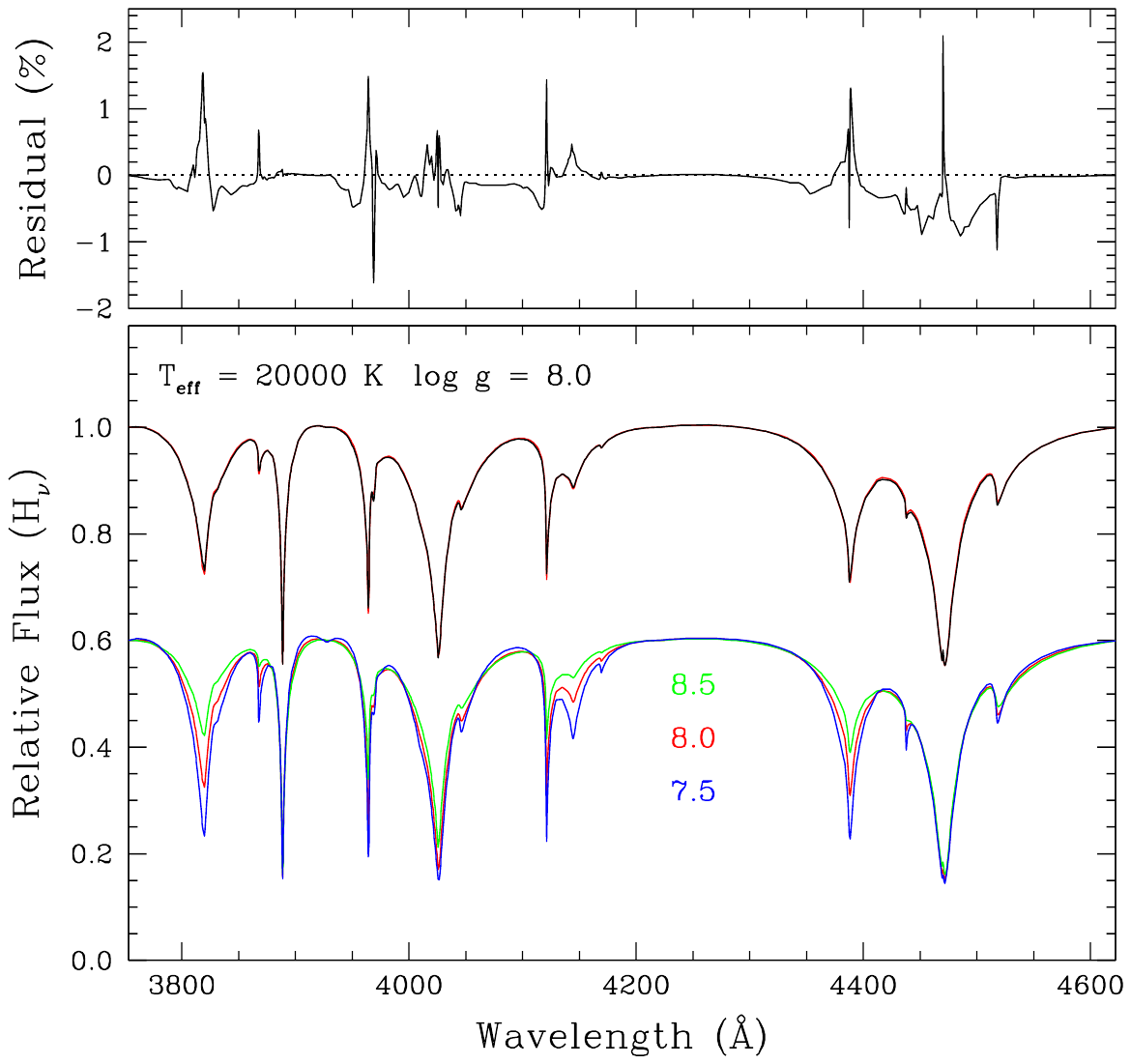}
\caption{Bottom panel: Comparison
  of normalized synthetic spectra (Eddington fluxes) at $\Te=20,000$~K and pure helium compositions.
  The top two spectra show the
  calculations at $\logg=8.0$ obtained with the B25 Stark profiles with no line
  dissolution (black) and the computer simulations from Paper~I (red),
  while the bottom spectra (shifted arbitrarily by 0.4) also rely on
  the Stark profiles from computer simulations but for different
  $\logg$ values indicated in the figure. All spectra are convolved
  with a 3~\AA\ (FWHM) Gaussian profile. Top panel: Residual fluxes of
  the $\logg=8$ normalized spectra.
\label{compsynth_PaperII}}
\end{figure}

In Figure~18 of Paper~I, we compared the absolute Eddington
fluxes of white dwarf model spectra at various effective temperatures
and $\logg=8$, using both sets of He~{\sc i} Stark profiles. Because
the spectroscopic method relies on normalized spectra rather than
absolute fluxes, we compare instead in Figure~\ref{compsynth_PaperII}
pure-helium DB model spectra at $\Te=20,000$~K, but normalized to a
continuum set to unity. The comparison is restricted to the wavelength
range where the He~{\sc i} lines are particularly sensitive to
$\logg$. The top two spectra in the bottom panel contrast the B25
Stark profiles at $\logg=8.0$ with no line dissolution (black) and the
computer simulations from Paper~I (red). Given these overlap almost
perfectly, we also show in the top panel the residual normalized
fluxes between these two spectra. The small differences, of order 1\%,
are likely related to the treatment of ion dynamics near the line
cores and to electron broadening between the line cores and the
continuum, which is handled in a significantly more sophisticated
manner in the simulation-based approach. Note that there are no
differences in the continuum regions because this is where the spectra
are normalized. Also shown in the bottom panel of
Figure~\ref{compsynth_PaperII} are model spectra computed with Stark
profiles from computer simulations but for different $\logg$
values. These illustrate how subtle differences in the overall line
profiles can affect the derived spectroscopic $\logg$ values.

\begin{figure}
% x1 y1 x2 y2
\includegraphics[clip=true,trim=1.2cm 9cm 1.2cm 2cm,width=\columnwidth]{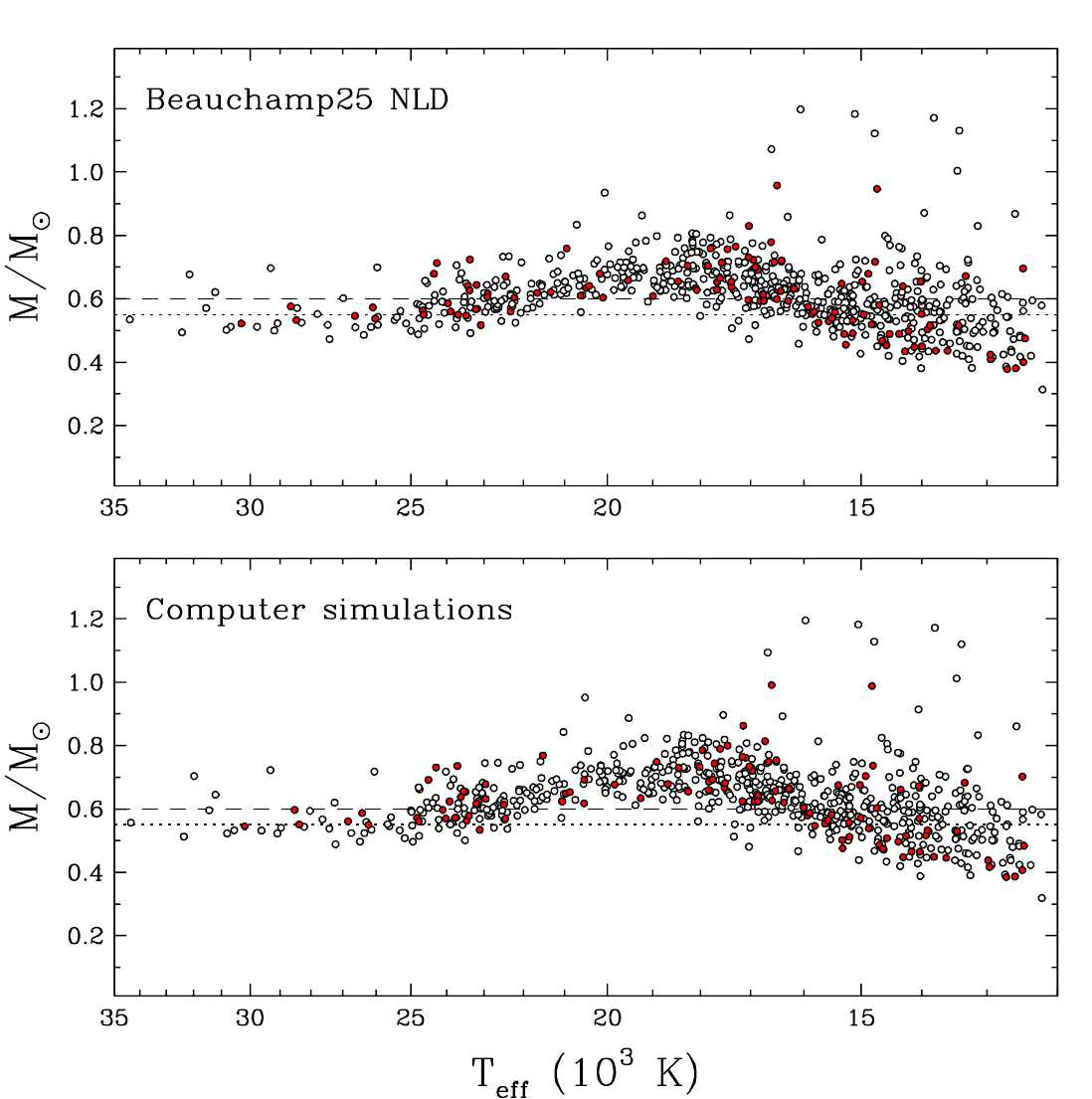}
\caption{Spectroscopic mass distributions as a function of $\Te$ for
  the DB white dwarfs in the DR17 SDSS sample (white circles) using
  the B25 Stark profiles and the computer simulations from Paper~I. In
  both sets of calculations, line dissolution is omitted (NLD), and
  van der Waals broadening is included using the Deridder \& van
  Rensbergen theory. Also shown in red are the results for the DB
  white dwarfs analyzed by \citet{Bergeron2011} and
  \citet{Rolland2018}. In both panels, the dashed and dotted lines
  correspond to constant masses of 0.6 and 0.55~\msun, respectively.
\label{correltm4}}
\end{figure}

We further compare in Figure~\ref{correltm4} the spectroscopic mass
distributions obtained from \hei\ Stark profiles based on our computer
simulations and the B25 semi-analytical calculations for the DB white
dwarfs in the SDSS sample. Also shown in red are updated results for
the DB white dwarfs analyzed by \citet{Bergeron2011} and
\citet{Rolland2018}, based on single-slit spectroscopic data (see next
paragraph). The differences between both sets of calculations are
barely noticeable in these diagrams, as anticipated from the results
shown in Figure~\ref{comptg4}. We still observe, however, that the
spectroscopic masses in the $\sim$18,000–21,000~K temperature range
remain overestimated compared with the canonical average mass of
0.6~\msun\ for white dwarfs. At lower temperatures, the masses are now
underestimated when using the van der Waals theory of Deridder \& van
Rensbergen. These masses can be increased, however, if the Unsöld
theory is adopted instead, an alternative approach explored in the
next subsection.

Finally, as mentioned in the Introduction, residual flux calibration
issues with the SDSS spectra have often been invoked to explain some
of the discrepancies in the measured spectroscopic parameters of both
DA and DB white dwarfs (see, e.g., \citealt{Genest2014};
\citealt{Genest2019b}). However, as shown in Figure~\ref{correltm4},
the results obtained for the DB white dwarfs analyzed by
\citet{Bergeron2011} and \citet{Rolland2018}, based on single-slit
spectroscopic data, now agree perfectly with the SDSS results. This
strongly suggests that the flux calibration problem affecting earlier
SDSS spectra has now been corrected. Consequently, flux calibration
issues can no longer be invoked to explain the discrepancies in the
spectroscopic mass distribution of DB white dwarfs.

\subsection{Broadening by Neutral Particles and 3D Hydrodynamical Corrections}\label{vdW}

Given that the Stark profiles from B97 have been widely used in the
spectroscopic analysis of DB white dwarfs, some of the results based
on these earlier calculations need to be revisited, especially for
cool DB stars.  As discussed above, we now rely exclusively on the B25
profile calculations with the proper inclusion of Doppler broadening
and normalization, with twice the frequency sampling (2X) of the
original tables of B97.  In this section, we now attempt to bring all
our results together in terms of \hei\ Stark profiles, broadening by
neutral particles, and 3D hydrodynamical corrections. For
completeness, we first begin by summarizing the two different He {\sc
  i} line broadening theories by neutral particles used in our
analysis.

In our models, we include only broadening by ground-state He {\sc i}
perturbers, which dominate cool DB atmospheres.  Resonance broadening
arises from a first-order resonant dipole-dipole interaction
\(V(r)\propto r^{-3}\) between identical atoms, and requires a
dipole-allowed transition between the ground state and either level of
the radiative transition.  Van der Waals broadening results from
second-order induced dipole-dipole interactions, and is always
present.  These definitions yield three cases for He {\sc i} line
broadening by neutral perturbers: (1) For different atomic species,
only van der Waals broadening applies and its contribution from
elements other than helium is negligible in DB/DBA stars; (2) For He
{\sc i} lines connecting triplet states (e.g., $\lambda$4471), no
dipole-allowed transition connects the ground state to either level of
the transition, so only van der Waals broadening occurs; (3) For
singlet He {\sc i} lines considered in this work (e.g.,
$\lambda$4922), resonance broadening is present because the lower or
upper level connects to the ground state, and the full interaction
includes both first- and second-order terms.

In our models, van der Waals and resonance broadening are combined
following the procedure described in detail by \citet[][see also
  \citealt{Genest2019a}]{BeauchampPhD}. Resonance broadening is
treated according to the theory of \cite{Ali1965}, while van der Waals
broadening is taken into account using the theory from
\cite{Unsold1955}, or an alternative approach we refer to as the
Deridder \& van Rensbergen formalism.  First, the width of the
Lorentzian profile ($\omega_{\rm vdW}$) is calculated twice: once
using the theory in \cite{Unsold1955} and again by applying the
approach described in \cite{Deridder1976}. The Deridder \& van
Rensbergen theory consistently predicts a broader profile when the
initial and final effective quantum numbers of the transition exceed 2
or 3. Because the Smirnov potential, which they used, is no longer
valid below these quantum numbers, we retain the following
conservative value for the width of the Lorentzian profile:
\begin{equation}
\omega_{\rm vdW}={\rm max}(\omega_{\rm Unsold},\omega_{\rm Deridder})\ .
\end{equation}
 
\noindent Empirically, \cite{BeauchampPhD} found that the neutral
helium lines in cool DB stars at $\lambda=4121$ \AA\ and 4713
\AA\ could be more accurately reproduced when treated strictly within
the \cite{Unsold1955} theory. This is the approach adopted here as
well.

Regarding transitions between singlet states, the available data in
the literature, to our knowledge, only provide separate estimates for
resonance and van der Waals broadening contributions. This raises the
issue of statistical dependence: it is not guaranteed that these two
mechanisms are independent, since a given collision may contribute to
both. Furthermore, assuming strong statistical dependance, it is not
clear whether the two contributions should add constructively or
destructively.  Rather than summing the two contributions to compute
the Lorentzian half-width—which would implicitly assume statistical
independence—we chose to retain the maximum of the two half-widths,
\begin{equation}
\omega_{\rm neutral}={\rm max}(\omega_{\rm resonance},\omega_{\rm vdW})
\end{equation}

\noindent rather than following the prescription by \cite{Lewis1967},
who determined that combining resonance and van der Waals broadening
resulted in a profile with $\omega_{\rm
  neutral}\sim0.6-0.8\,(\omega_{\rm resonance}+\omega_{\rm
  vdW})$. However, that study was limited to temperatures around 100
K, which are not applicable to our DB models.

\begin{figure}
% x1 y1 x2 y2
\includegraphics[clip=true,trim=1.2cm 1cm 1.2cm 2cm,width=\columnwidth]{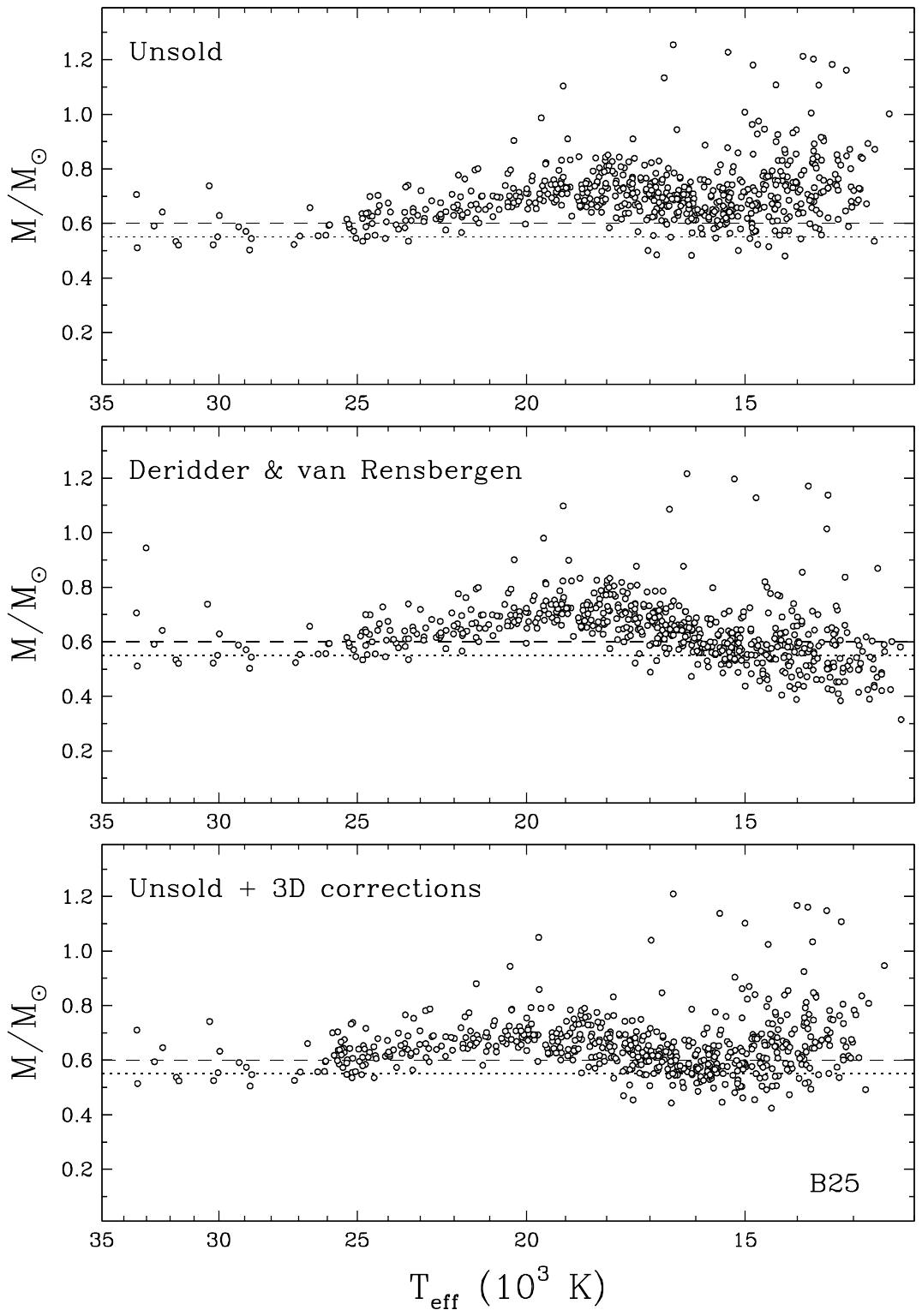}
\caption{Spectroscopic mass distributions as a function of $\Te$ for
the DB white dwarfs in the DR17 SDSS sample using the B25
Stark profiles, including line dissolution. The top and middle
panels differ in terms of the theory used for van der Waals
broadening, as indicated in each panel, while the bottom panel is
similar to the top panel but with the 3D hydrodynamical corrections
of \citet{Cukanovaite2021} taken into account. In all panels, the
dashed and dotted lines correspond to constant masses of 0.6 and
0.55~\msun, respectively.
\label{correltm7}}
\end{figure}

We first show in Figure~\ref{correltm7} the spectroscopic mass
distributions as a function of $\Te$ for our DB white dwarf SDSS
sample using the B25 Stark profiles, which, for the purpose of this
discussion, now include line dissolution. These calculations thus
represent the state-of-the-art Stark profiles within the
semi-analytical framework of B97. The top two panels rely on the
different treatments of van der Waals broadening described above,
namely the \citet[][top panel]{Unsold1955} theory and the Deridder \&
van Rensbergen formalism (middle panel), while the bottom panel also
adopts the Unsöld theory but includes the 3D hydrodynamical
corrections of \citet{Cukanovaite2021}.

The effect of our corrected treatment of Doppler broadening on the mass
distribution can be appreciated by comparing the middle panel of
Figure \ref{correltm7} with the bottom panel of Figure
\ref{correltm1}, which relies on the same van der Waals broadening
theory but differs in terms of the way Doppler broadening is taken
into acount; note that for clarity, DB stars with weak \hei\ lines
(${\rm EW}_{4471}< 5$ \AA) are not shown in Figure
\ref{correltm7}.
While our earlier
semi-analytical profiles with inaccurate Doppler broadening favored
the Deridder \& van Rensbergen treatment of van der Waals broadening,
our revised B25 profiles
used in the middle panel of Figure~\ref{correltm7} clearly show that this approach now
yields masses that are too low below $\Te\sim17,000$~K. The Unsöld theory (top panel), on the other hand,
produces masses that appear slightly overestimated. Given that 3D
hydrodynamical simulations tend to lower the inferred $\logg$ values
compared to those derived from 1D model atmospheres (see Figure~5 of
\citealt{Cukanovaite2021}), we can significantly reduce the
spectroscopic masses by applying the 3D hydrodynamical corrections to
the Unsöld determinations, as shown in the bottom panel of
Figure~\ref{correltm7}. The resulting mass distribution below
$\Te\sim17,000$~K now follows the canonical 0.6~\msun\ mean mass much
more closely — in better agreement with the photometric mass
distribution displayed in the top panel of
Figure~\ref{correltm1}. However, a clear discrepancy remains in the
range $24,000~{\rm K} > \Te > 17,000$~K, where the masses are still
predicted to be too high, even after including the 3D corrections.

\begin{figure}
% x1 y1 x2 y2
\includegraphics[clip=true,trim=1.2cm 1cm 1.2cm 2cm,width=\columnwidth]{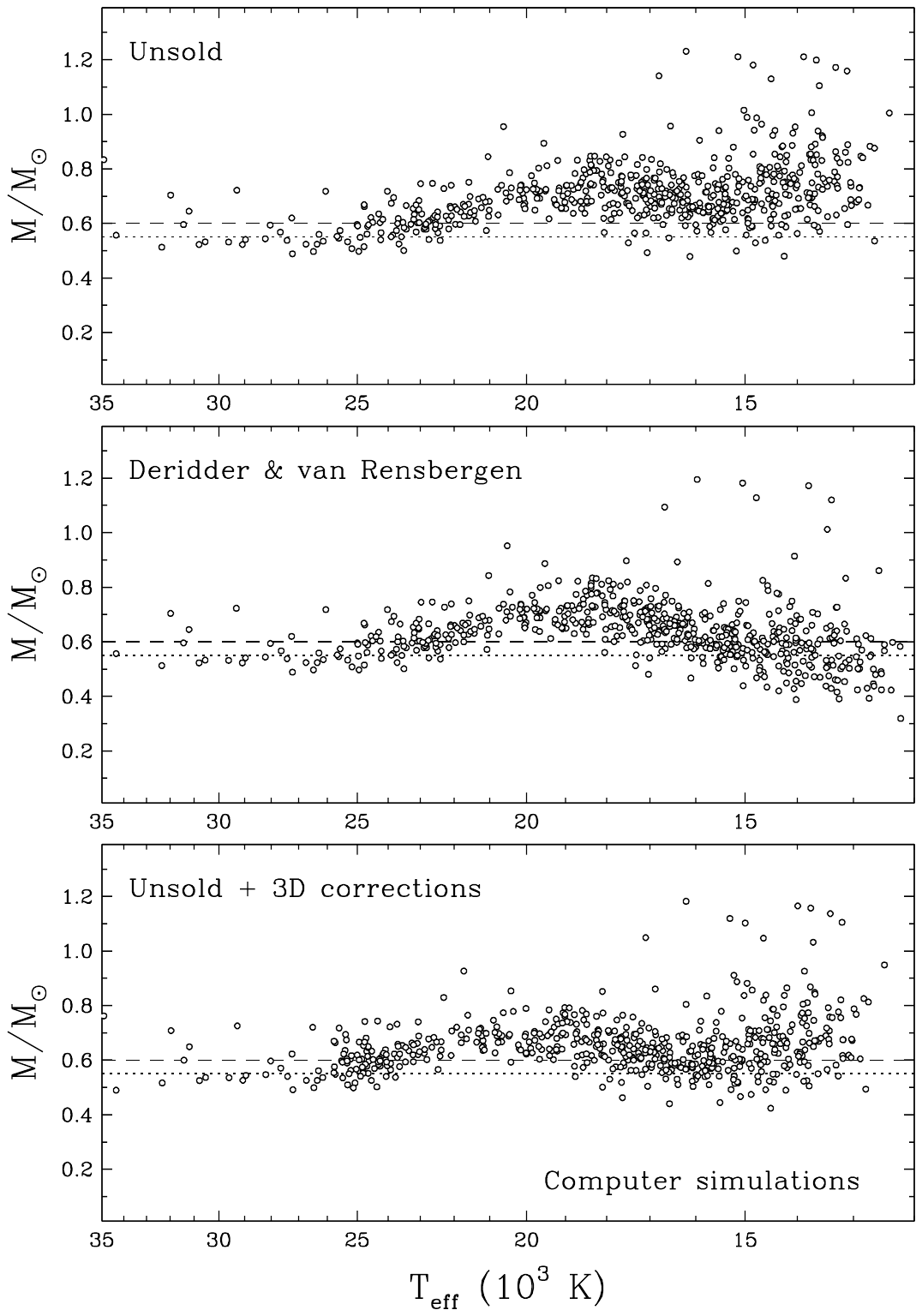}
\caption{Same as Figure~\ref{correltm7} but using the \hei\ Stark profiles
from the computer simulations of Paper~I.
\label{correltm5}}
\end{figure}

We now turn to the results shown in Figure~\ref{correltm5}, obtained
under assumptions similar to those in Figure~\ref{correltm7}, but this
time using the \hei\ Stark profiles computed from our computer
simulations. We remind the reader that line dissolution is not
included in these calculations. This omission largely explains why the
spectroscopic masses appear lower at higher temperatures. As already
demonstrated in Figure~\ref{comptg3}, neglecting line dissolution
decreases the $\logg$ values by roughly 0.05~dex (or about
0.03~\msun\ in mass) for hot ($\Te>15,000$~K) DB white dwarfs.

The results displayed in Figure~\ref{correltm5} obtained from computer
simulations are practically equivalent to those shown in
Figure~\ref{correltm7}, based on the semi-analytical B25
calculations. We are thus forced to conclude that our significant
improvements in the microphysics of the helium atom have not resolved
the long-standing problem with the spectroscopic mass distribution of
DB white dwarfs originally identified by
\citet{Genest2019b}. Furthermore, the inclusion of 3D hydrodynamical
corrections fails to reduce the spectroscopic masses in the
$24,000~{\rm K} > \Te > 17,000$~K range, and the discrepancy remains.

\section{Concluding Remarks}\label{conclusion}

We presented in this paper improved Stark-broadening calculations for
neutral helium lines based on state-of-the-art computer simulations
that incorporate corrections for ion dynamics, transitions of electron
contributions from the core to the wings of the profile, numerical
integration of the time-evolution operator for helium perturbed by a
fluctuating electric field, the Debye correction for charge–motion
correlations, local density variations, and particle reinjection. From
a theoretical standpoint, these line profiles represent a major
advance over the semi-analytical B97 profiles used in all previous
studies. In those earlier calculations, electron broadening was
treated using the impact approximation near the line core and the
one-electron approximation in the far wings, with an interpolation
scheme applied in an arbitrary fashion to ensure a smooth transition
between the two regimes. In our simulations, this transition arises
naturally from the underlying physics.

A key improvement in our approach is the explicit inclusion of ion
dynamics, an effect completely neglected in the semi-analytical
method. While ion dynamics has no detectable impact in the context of
white dwarfs, it plays an important role in lower-gravity stars, as
demonstrated in Figures~20 and 21 of Paper~I, where the limitations
of the quasi-static approximation are clearly illustrated.

In \citet{TremblayP2020}, we introduced new computer-simulation
calculations of Stark-broadened He~{\sc i} profiles that unified the
treatment of ions and electrons, but severe numerical issues prevented
us from presenting extensive tables. These problems have now been
overcome thanks to the new numerical procedure described in detail in
Paper~I. The central improvement is the use of the power-spectrum
formalism for line-profile generation, rather than the autocorrelation
function used previously. As a result, detailed calculations are now
available for the most important optical He~{\sc i} lines, over a wide
range of temperatures and electron densities.

A major motivation behind this project was to address the discrepancy
illustrated in Figure~\ref{correltm1}: the photometric mass
distribution remains centered near the canonical value of 0.6 \msun,
whereas the spectroscopic mass distribution deviates significantly
from this value over several temperature ranges—even after excluding
the coolest objects (shown in red) whose very weak helium lines
(He~{\sc i} $\lambda$4471 equivalent width smaller than 5~\AA) render
spectroscopic determinations unreliable. These comparisons point to a
fundamental issue with the spectroscopic technique, for which detailed
He~{\sc i} Stark profiles are a crucial ingredient. Our final results,
summarized in Figures~\ref{correltm7} and \ref{correltm5}, indicate
that the discrepancy persists, most notably in the range $24,000~{\rm
  K} > \Te > 17,000~{\rm K}$, regardless of the Stark-broadening
theory adopted, the treatment of van der Waals broadening at low
temperatures, or the inclusion of 3D hydrodynamical effects. We also
demonstrated that the use of SDSS spectra—known to be problematic in
earlier work—is not responsible for the discrepancy. A different
explanation must therefore be sought.

Given that the results shown in Figures~\ref{correltm7} and
\ref{correltm5} are quite comparable (see also Figure~\ref{comptg4}),
it is likely preferable, for the purpose of atmospheric analyses of
white dwarfs, to continue relying on our semi-analytical Stark
profiles, which incorporate line dissolution—a non-negligible effect
at high temperatures, as illustrated in Figure~\ref{comptg3}. The
impact on spectroscopic masses can be inferred by comparing
Figures~\ref{correltm7} and \ref{correltm5}. Although the results of
our computer simulations may appear discouraging in this context, it
is important to recognize that they have enabled substantial
improvements to our semi-analytical profiles. Chief among these is our
revised treatment of Doppler broadening, which has significant
consequences at the cool end of the DB sequence (see
Figure~\ref{comptg1}), where the treatment of broadening by
  neutral particles—still a poorly understood mechanism—also plays an
important role (see Figure~\ref{correltm7}). Our simulations
additionally revealed several issues in the semi-analytical
calculations, including improper frequency sampling and
normalization. Finally, they may eventually lead to a more physically
motivated interpolation scheme between the one-electron and impact
regimes, as briefly discussed in Paper~I.

Despite the progress presented here, several improvements to the
Stark-broadened He~{\sc i} profiles are still possible. For instance,
while staying in the semi-classical approximation: (1) highest order
terms in the multipole expansion of the potential, (2) mixing of upper
states with different principal quantum number $n$, and of course, (3)
quantum effects such as penetrating collisions
\citep{Gomez2017,Cho2022}. However, such refinements are probably
second-order and are unlikely to resolve the persistent spectroscopic
mass discrepancy for DB white dwarfs in the 17,000–24,000~K
temperature range. Likewise, significant further
improvements from 3D hydrodynamical models are improbable. Other
explanations must be considered, possibly involving the temperature
and pressure structures themselves—such as missing opacity sources, or
known sources that remain poorly modeled, including resonance lines in
the UV (see Chapter~6 of \citealt{BeauchampPhD}) or the pseudo-continuum
opacity associated with dissolved levels of helium, a process usually
neglected for the ground state (in hydrogen as well).

On a more encouraging note, our corrected treatment of Doppler
broadening substantially mitigates the high-$\log g$ problem commonly
reported at the cool end of the DB sequence (see
Figures~\ref{correltm7} and \ref{correltm5}), although progress in
modeling the broadening by neutral particles remains urgently needed.

\begin{acknowledgements}
  We thank A. B{\'e}dard for a careful reading of our manuscript and
  his constructive comments, and O.~Vincent for helping us with the
  extraction of the relevant SDSS photometric and spectroscopic data
  used in this analysis.  This work was supported in part by the NSERC
  Canada and by the Fund FRQNT (Qu\'ebec).

\end{acknowledgements}

\section*{Data availability}
Both the semi-analytical (with and without line dissolution) and
simulation He {\sc i} line profiles are available on Zenodo, which can
be accessed via the following link
\dataset[doi:10.5281/zenodo.18722143]{https://doi.org/10.5281/zenodo.18722143}.

\bibliographystyle{aasjournal}
\bibliography{ms}

@ARTICLE{Bergeron1992,
       author = {{Bergeron}, P. and {Saffer}, Rex A. and {Liebert}, James},
        title = "{A Spectroscopic Determination of the Mass Distribution of DA White Dwarfs}",
      journal = {\apj},
         year = 1992,
        month = jul,
       volume = {394},
        pages = {228},
          doi = {10.1086/171575},
       adsurl = {https://ui.adsabs.harvard.edu/abs/1992ApJ...394..228B}
}

@ARTICLE{GentileFusillo2019,
   author = {{Gentile Fusillo}, N.~P. and {Tremblay}, P.-E. and {G{\"a}nsicke}, B.~T. and 
	{Manser}, C.~J. and {Cunningham}, T. and {Cukanovaite}, E. and 
	{Hollands}, M. and {Marsh}, T. and {Raddi}, R. and {Jordan}, S. and 
	{Toonen}, S. and {Geier}, S. and {Barstow}, M. and {Cummings}, J.~D.
	},
    title = "{A Gaia Data Release 2 catalogue of white dwarfs and a comparison with SDSS}",
  journal = {\mnras},
archivePrefix = "arXiv",
   eprint = {1807.03315},
 primaryClass = "astro-ph.SR",
     year = 2019,
    month = feb,
   volume = 482,
    pages = {4570-4591},
      doi = {10.1093/mnras/sty3016},
   adsurl = {http://adsabs.harvard.edu/abs/2019MNRAS.482.4570G}
}

@ARTICLE{Fontaine2001,
   author = {{Fontaine}, G. and {Brassard}, P. and {Bergeron}, P.},
    title = "{The Potential of White Dwarf Cosmochronology}",
  journal = {\pasp},
     year = 2001,
    month = apr,
   volume = 113,
    pages = {409-435},
      doi = {10.1086/319535},
   adsurl = {http://adsabs.harvard.edu/abs/2001PASP..113..409F}
}

@ARTICLE{Bergeron2011,
   author = {{Bergeron}, P. and {Wesemael}, F. and {Dufour}, P. and {Beauchamp}, A. and 
	{Hunter}, C. and {Saffer}, R.~A. and {Gianninas}, A. and {Ruiz}, M.~T. and 
	{Limoges}, M.-M. and {Dufour}, P. and {Fontaine}, G. and {Liebert}, J.
	},
    title = "{A Comprehensive Spectroscopic Analysis of DB White Dwarfs}",
  journal = {\apj},
archivePrefix = "arXiv",
   eprint = {1105.5433},
 primaryClass = "astro-ph.SR",
     year = 2011,
    month = aug,
   volume = 737,
      eid = {28},
    pages = {28},
      doi = {10.1088/0004-637X/737/1/28},
   adsurl = {http://adsabs.harvard.edu/abs/2011ApJ...737...28B}
}

@ARTICLE{Liebert2005,
   author = {{Liebert}, J. and {Bergeron}, P. and {Holberg}, J.~B.},
    title = "{The Formation Rate and Mass and Luminosity Functions of DA White Dwarfs from the Palomar Green Survey}",
  journal = {\apjs},
   eprint = {astro-ph/0406657},
     year = 2005,
    month = jan,
   volume = 156,
    pages = {47-68},
      doi = {10.1086/425738},
   adsurl = {http://adsabs.harvard.edu/abs/2005ApJS..156...47L}
}

@ARTICLE{Gaia2018,
   author = {{Gaia Collaboration} and {Babusiaux}, C. and {van Leeuwen}, F. and 
	{Barstow}, M.~A. and {Jordi}, C. and {Vallenari}, A. and {Bossini}, D. and 
	{Bressan}, A. and {Cantat-Gaudin}, T. and {van Leeuwen}, M. and et al.},
    title = "{Gaia Data Release 2. Observational Hertzsprung-Russell diagrams}",
  journal = {\aap},
archivePrefix = "arXiv",
   eprint = {1804.09378},
 primaryClass = "astro-ph.SR",
     year = 2018,
    month = aug,
   volume = 616,
      eid = {A10},
    pages = {A10},
      doi = {10.1051/0004-6361/201832843},
   adsurl = {http://adsabs.harvard.edu/abs/2018A%26A...616A..10G}
}

@ARTICLE{Tremblay2019,
       author = {{Tremblay}, P. -E. and {Cukanovaite}, E. and {Gentile Fusillo}, N.~P.
        and {Cunningham}, T. and {Hollands}, M.~A.},
        title = "{Fundamental parameter accuracy of DA and DB white dwarfs in Gaia Data
        Release 2}",
      journal = {\mnras},
         year = 2019,
        month = Feb,
       volume = {482},
        pages = {5222-5232},
          doi = {10.1093/mnras/sty3067},
archivePrefix = {arXiv},
       eprint = {1811.03084},
 primaryClass = {astro-ph.SR},
       adsurl = {https://ui.adsabs.harvard.edu/#abs/2019MNRAS.482.5222T}
}

@ARTICLE{Rolland2018,
       author = {{Rolland}, B. and {Bergeron}, P. and {Fontaine}, G.},
        title = "{On the Spectral Evolution of Helium-atmosphere White Dwarfs Showing Traces of Hydrogen}",
      journal = {\apj},
         year = 2018,
        month = apr,
       volume = {857},
       number = {1},
          eid = {56},
        pages = {56},
          doi = {10.3847/1538-4357/aab713},
archivePrefix = {arXiv},
       eprint = {1803.05965},
 primaryClass = {astro-ph.SR},
       adsurl = {https://ui.adsabs.harvard.edu/abs/2018ApJ...857...56R}
}

@ARTICLE{Tremblay2011,
   author = {{Tremblay}, P.-E. and {Bergeron}, P. and {Gianninas}, A.},
    title = "{An Improved Spectroscopic Analysis of DA White Dwarfs from the Sloan Digital Sky Survey Data Release 4}",
  journal = {\apj},
archivePrefix = "arXiv",
   eprint = {1102.0056},
 primaryClass = "astro-ph.SR",
     year = 2011,
    month = apr,
   volume = 730,
      eid = {128},
    pages = {128},
      doi = {10.1088/0004-637X/730/2/128},
   adsurl = {http://adsabs.harvard.edu/abs/2011ApJ...730..128T}
}

@ARTICLE{Bedard2020,
       author = {{B{\'e}dard}, A. and {Bergeron}, P. and {Brassard}, P. and {Fontaine}, G.},
        title = "{On the Spectral Evolution of Hot White Dwarf Stars. I. A Detailed Model Atmosphere Analysis of Hot White Dwarfs from SDSS DR12}",
      journal = {\apj},
         year = 2020,
        month = oct,
       volume = {901},
       number = {2},
          eid = {93},
        pages = {93},
          doi = {10.3847/1538-4357/abafbe},
archivePrefix = {arXiv},
       eprint = {2008.07469},
 primaryClass = {astro-ph.SR},
       adsurl = {https://ui.adsabs.harvard.edu/abs/2020ApJ...901...93B}
}

@ARTICLE{Genest2019a,
       author = {{Genest-Beaulieu}, C. and {Bergeron}, P.},
        title = "{A Comprehensive Spectroscopic and Photometric Analysis of DA and DB White Dwarfs from SDSS and Gaia}",
      journal = {\apj},
         year = 2019,
        month = feb,
       volume = {871},
       number = {2},
          eid = {169},
        pages = {169},
          doi = {10.3847/1538-4357/aafac6},
       adsurl = {https://ui.adsabs.harvard.edu/abs/2019ApJ...871..169G}
}

@ARTICLE{Genest2019b,
       author = {{Genest-Beaulieu}, C. and {Bergeron}, P.},
        title = "{A Photometric and Spectroscopic Investigation of the DB White Dwarf Population Using SDSS and Gaia Data}",
      journal = {\apj},
         year = 2019,
        month = sep,
       volume = {882},
       number = {2},
          eid = {106},
        pages = {106},
          doi = {10.3847/1538-4357/ab379e},
archivePrefix = {arXiv},
       eprint = {1908.01728},
 primaryClass = {astro-ph.SR},
       adsurl = {https://ui.adsabs.harvard.edu/abs/2019ApJ...882..106G}
}

@ARTICLE{Bergeron1997,
   author = {{Bergeron}, P. and {Ruiz}, M.~T. and {Leggett}, S.~K.},
    title = "{The Chemical Evolution of Cool White Dwarfs and the Age of the Local Galactic Disk}",
  journal = {\apjs},
     year = 1997,
    month = jan,
   volume = 108,
    pages = {339-387},
      doi = {10.1086/312955},
   adsurl = {http://adsabs.harvard.edu/abs/1997ApJS..108..339B}
}

@ARTICLE{Bergeron2019,
       author = {{Bergeron}, P. and {Dufour}, P. and {Fontaine}, G. and {Coutu}, S. and {Blouin}, S. and {Genest-Beaulieu}, C. and {B{\'e}dard}, A. and {Rolland}, B.},
        title = "{On the Measurement of Fundamental Parameters of White Dwarfs in the Gaia Era}",
      journal = {\apj},
         year = 2019,
        month = may,
       volume = {876},
       number = {1},
          eid = {67},
        pages = {67},
          doi = {10.3847/1538-4357/ab153a},
archivePrefix = {arXiv},
       eprint = {1904.02022},
 primaryClass = {astro-ph.SR},
       adsurl = {https://ui.adsabs.harvard.edu/abs/2019ApJ...876...67B}
}

@ARTICLE{Cukanovaite2021,
       author = {{Cukanovaite}, Elena and {Tremblay}, Pier-Emmanuel and {Bergeron}, Pierre and {Freytag}, Bernd and {Ludwig}, Hans-G{\"u}nter and {Steffen}, Matthias},
        title = "{3D spectroscopic analysis of helium-line white dwarfs}",
      journal = {\mnras},
         year = 2021,
        month = mar,
       volume = {501},
       number = {4},
        pages = {5274-5293},
          doi = {10.1093/mnras/staa3684},
archivePrefix = {arXiv},
       eprint = {2011.12693},
 primaryClass = {astro-ph.SR},
       adsurl = {https://ui.adsabs.harvard.edu/abs/2021MNRAS.501.5274C}
}

@ARTICLE{Kilic2020,
       author = {{Kilic}, Mukremin and {Bergeron}, P. and {Kosakowski}, Alekzander and {Brown}, Warren R. and {Ag{\"u}eros}, Marcel A. and {Blouin}, Simon},
        title = "{The 100 pc White Dwarf Sample in the SDSS Footprint}",
      journal = {\apj},
         year = 2020,
        month = jul,
       volume = {898},
       number = {1},
          eid = {84},
        pages = {84},
          doi = {10.3847/1538-4357/ab9b8d},
archivePrefix = {arXiv},
       eprint = {2006.00323},
 primaryClass = {astro-ph.SR},
       adsurl = {https://ui.adsabs.harvard.edu/abs/2020ApJ...898...84K}
}

@ARTICLE{Bergeron2001,
       author = {{Bergeron}, P. and {Leggett}, S.~K. and {Ruiz}, Mar{\'\i}a Teresa},
        title = "{Photometric and Spectroscopic Analysis of Cool White Dwarfs with Trigonometric Parallax Measurements}",
      journal = {\apjs},
         year = 2001,
        month = apr,
       volume = {133},
       number = {2},
        pages = {413-449},
          doi = {10.1086/320356},
archivePrefix = {arXiv},
       eprint = {astro-ph/0011286},
 primaryClass = {astro-ph},
       adsurl = {https://ui.adsabs.harvard.edu/abs/2001ApJS..133..413B}
}

@ARTICLE{Genest2014,
   author = {{Genest-Beaulieu}, C. and {Bergeron}, P.},
    title = "{Comparison of Atmospheric Parameters Determined from Spectroscopy and Photometry for DA White Dwarfs in the Sloan Digital Sky Survey}",
  journal = {\apj},
archivePrefix = "arXiv",
   eprint = {1410.5255},
 primaryClass = "astro-ph.SR",
     year = 2014,
    month = dec,
   volume = 796,
      eid = {128},
    pages = {128},
      doi = {10.1088/0004-637X/796/2/128},
   adsurl = {http://adsabs.harvard.edu/abs/2014ApJ...796..128G}
}

@ARTICLE{Cukanovaite2018,
       author = {{Cukanovaite}, E. and {Tremblay}, P. -E. and {Freytag}, B. and {Ludwig},
        H. -G. and {Bergeron}, P.},
        title = "{Pure-helium 3D model atmospheres of white dwarfs}",
      journal = {\mnras},
         year = 2018,
        month = Dec,
       volume = {481},
        pages = {1522-1537},
          doi = {10.1093/mnras/sty2383},
archivePrefix = {arXiv},
       eprint = {1809.00590},
 primaryClass = {astro-ph.SR},
       adsurl = {https://ui.adsabs.harvard.edu/#abs/2018MNRAS.481.1522C}
}

@ARTICLE{TremblayP2020,
       author = {{Tremblay}, Patrick and {Beauchamp}, A. and {Bergeron}, P.},
        title = "{New Calculations of Stark-broadened Profiles for Neutral Helium Lines Using Computer Simulations}",
      journal = {\apj},
         year = 2020,
        month = oct,
       volume = {901},
       number = {2},
          eid = {104},
        pages = {104},
          doi = {10.3847/1538-4357/abb0e5},
archivePrefix = {arXiv},
       eprint = {2008.09834},
 primaryClass = {astro-ph.SR},
       adsurl = {https://ui.adsabs.harvard.edu/abs/2020ApJ...901..104T}
}

@ARTICLE{beauchamp97,
       author = {{Beauchamp}, A. and {Wesemael}, F. and {Bergeron}, P.},
        title = "{Spectroscopic Studies of DB White Dwarfs: Improved Stark Profiles for Optical Transitions of Neutral Helium}",
      journal = {\apjs},
         year = "1997",
        month = "Feb",
       volume = {108},
       number = {2},
        pages = {559-573},
          doi = {10.1086/312961},
       adsurl = {https://ui.adsabs.harvard.edu/abs/1997ApJS..108..559B}
}

@ARTICLE{Seaton1990,
       author = {{Seaton}, M.~J.},
        title = "{Atomic data for opacity calculations. XIII. Line profiles for transitions in hydrogenic ions}",
      journal = {Journal of Physics B Atomic Molecular Physics},
         year = 1990,
        month = oct,
       volume = {23},
       number = {19},
        pages = {3255-3296},
          doi = {10.1088/0953-4075/23/19/012},
       adsurl = {https://ui.adsabs.harvard.edu/abs/1990JPhB...23.3255S}
}

@ARTICLE{HM1988,
       author = {{Hummer}, D.~G. and {Mihalas}, Dimitri},
        title = "{The Equation of State for Stellar Envelopes. I. an Occupation Probability Formalism for the Truncation of Internal Partition Functions}",
      journal = {\apj},
         year = 1988,
        month = aug,
       volume = {331},
        pages = {794},
          doi = {10.1086/166600},
       adsurl = {https://ui.adsabs.harvard.edu/abs/1988ApJ...331..794H}
}

@ARTICLE{Tremblay2009,
       author = {{Tremblay}, P. -E. and {Bergeron}, P.},
        title = "{Spectroscopic Analysis of DA White Dwarfs: Stark Broadening of Hydrogen Lines Including Nonideal Effects}",
      journal = {\apj},
         year = 2009,
        month = may,
       volume = {696},
       number = {2},
        pages = {1755-1770},
          doi = {10.1088/0004-637X/696/2/1755},
archivePrefix = {arXiv},
       eprint = {0902.4182},
 primaryClass = {astro-ph.SR},
       adsurl = {https://ui.adsabs.harvard.edu/abs/2009ApJ...696.1755T}
}

@BOOK{Unsold1955,
       author = {{Unsold}, Albrecht},
        title = "{Physik der Sternatmospharen, MIT besonderer Berucksichtigung der Sonne.}",
    booktitle = {Berlin, Springer, 1955. 2. Aufl.},
    publisher = {Springer},
       volume = 2,
         year = 1955,
       adsurl = {https://ui.adsabs.harvard.edu/#abs/1955psmb.book.....U}
}

@ARTICLE{Deridder1976,
   author = {{Deridder}, G. and {van Rensbergen}, W.},
    title = "{Tables of damping constants of spectral lines broadened by H and He}",
  journal = {\aaps},
     year = 1976,
    month = feb,
   volume = 23,
    pages = {147},
   adsurl = {http://adsabs.harvard.edu/abs/1976A%26AS...23..147D}
}

@ARTICLE{Lallement2014,
       author = {{Lallement}, R. and {Vergely}, J. -L. and {Valette}, B. and {Puspitarini}, L. and {Eyer}, L. and {Casagrande}, L.},
        title = "{3D maps of the local ISM from inversion of individual color excess measurements}",
      journal = {\aap},
         year = 2014,
        month = jan,
       volume = {561},
          eid = {A91},
        pages = {A91},
          doi = {10.1051/0004-6361/201322032},
archivePrefix = {arXiv},
       eprint = {1309.6100},
 primaryClass = {astro-ph.GA},
       adsurl = {https://ui.adsabs.harvard.edu/abs/2014A&A...561A..91L}
}

@ARTICLE{Capitanio2017,
       author = {{Capitanio}, L. and {Lallement}, R. and {Vergely}, J.~L. and {Elyajouri}, M. and {Monreal-Ibero}, A.},
        title = "{Three-dimensional mapping of the local interstellar medium with composite data}",
      journal = {\aap},
         year = 2017,
        month = oct,
       volume = {606},
          eid = {A65},
        pages = {A65},
          doi = {10.1051/0004-6361/201730831},
archivePrefix = {arXiv},
       eprint = {1706.07711},
 primaryClass = {astro-ph.GA},
       adsurl = {https://ui.adsabs.harvard.edu/abs/2017A&A...606A..65C}
}

@INPROCEEDINGS{Lewis1967,
   author = {{Lewis}, E. L.},
    title = "{Self-broadening and oscillator strengths in the rare gases}",
booktitle = {Proc. Phys. Soc.},
     year = 1967,
   volume = 92,
    pages = {817},
}

@phdthesis{BeauchampPhD,
  author       = {Alain Beauchamp}, 
  title        = "{Détermination des paramètres atmosphériques des étoiles naines blanches de type DB}",
  school       = {Universit\'e de Montr\'eal},
  year         = 1995,
  month        = apr
}

@article{Ali1965,
  title = {Theory of Resonance Broadening of Spectral Lines by Atom-Atom
Impacts},
  author = {Ali, A. W. and Griem, H. R.},
  journal = {Phys. Rev.},
  volume = {140},
  issue = {4A},
  pages = {A1044--A1049},
  numpages = {0},
  year = {1965},
  month = {Nov},
  publisher = {American Physical Society},
  doi = {10.1103/PhysRev.140.A1044},
}

@ARTICLE{Griem62,
       author = {{Griem}, H.~R. and {Baranger}, M. and {Kolb}, A.~C. and {Oertel}, G.},
        title = "{Stark Broadening of Neutral Helium Lines in a Plasma}",
      journal = {Physical Review},
         year = 1962,
        month = jan,
       volume = {125},
       number = {1},
        pages = {177-195},
          doi = {10.1103/PhysRev.125.177},
       adsurl = {https://ui.adsabs.harvard.edu/abs/1962PhRv..125..177G}
}

@ARTICLE{BCS69,
       author = {{Barnard}, A.~J. and {Cooper}, J. and {Shamey}, L.~J.},
        title = "{Calculated Profiles of He I 4471 and 4922 A and their Forbidden Components}",
      journal = {\aap},
         year = "1969",
        month = "Jan",
       volume = {1},
        pages = {28},
       adsurl = {https://ui.adsabs.harvard.edu/abs/1969A&A.....1...28B}
}

@ARTICLE{BC70,
       author = {{Barnard}, A.~J. and {Cooper}, J.},
        title = "{Computed profiles of He I 5016 {\r{A}} at high electron densities.}",
      journal = {\jqsrt},
         year = "1970",
        month = "Jan",
       volume = {10},
       number = {6},
        pages = {695-702},
          doi = {10.1016/0022-4073(70)90085-3},
       adsurl = {https://ui.adsabs.harvard.edu/abs/1970JQSRT..10..695B}
}

@ARTICLE{Gig96,
       author = {{Gigosos}, Marco A. and {Carde{\~n}oso}, Valent{\'\i}n},
        title = "{New plasma diagnosis tables of hydrogen Stark broadening including ion dynamics}",
      journal = {Journal of Physics B Atomic Molecular Physics},
         year = "1996",
        month = "Oct",
       volume = {29},
       number = {20},
        pages = {4795-4838},
          doi = {10.1088/0953-4075/29/20/029},
       adsurl = {https://ui.adsabs.harvard.edu/abs/1996JPhB...29.4795G}
}

@ARTICLE{Gig09,
       author = {{Gigosos}, M.~A. and {Gonz{\'a}lez}, M. {\'A}.},
        title = "{Stark broadening tables for the helium I 447.1 line. Application to weakly coupled plasmas diagnostics}",
      journal = {\aap},
         year = "2009",
        month = "Aug",
       volume = {503},
       number = {1},
        pages = {293-299},
          doi = {10.1051/0004-6361/200912243},
       adsurl = {https://ui.adsabs.harvard.edu/abs/2009A&A...503..293G}
}

@ARTICLE{Gig85,
       author = {{Gigosos}, M.~A. and {Fraile}, J. and {Torres}, F.},
        title = "{Hydrogen Stark profiles: A simulation-oriented mathematical simplification}",
      journal = {\pra},
         year = "1985",
        month = "May",
       volume = {31},
       number = {5},
        pages = {3509-3511},
          doi = {10.1103/PhysRevA.31.3509},
       adsurl = {https://ui.adsabs.harvard.edu/abs/1985PhRvA..31.3509G}
}

@ARTICLE{Gig87,
       author = {{Gigosos}, M.~A. and {Cardenoso}, V.},
        title = "{Study of the effects of ion dynamics on Stark profiles of Balmer-{\ensuremath{\alpha}} and -{\ensuremath{\beta}} lines using simulation techniques}",
      journal = {Journal of Physics B Atomic Molecular Physics},
         year = "1987",
        month = "Nov",
       volume = {20},
       number = {22},
        pages = {6005-6019},
          doi = {10.1088/0022-3700/20/22/013},
       adsurl = {https://ui.adsabs.harvard.edu/abs/1987JPhB...20.6005G}
}

@ARTICLE{Finley1997,
       author = {{Finley}, David S. and {Koester}, Detlev and {Basri}, Gibor},
        title = "{The Temperature Scale and Mass Distribution of Hot DA White Dwarfs}",
      journal = {\apj},
         year = 1997,
        month = oct,
       volume = {488},
       number = {1},
        pages = {375-396},
          doi = {10.1086/304668},
       adsurl = {https://ui.adsabs.harvard.edu/abs/1997ApJ...488..375F}
}

@ARTICLE{Holberg2006,
       author = {{Holberg}, J.~B. and {Bergeron}, Pierre},
        title = "{Calibration of Synthetic Photometry Using DA White Dwarfs}",
      journal = {\aj},
         year = 2006,
        month = sep,
       volume = {132},
       number = {3},
        pages = {1221-1233},
          doi = {10.1086/505938},
       adsurl = {https://ui.adsabs.harvard.edu/abs/2006AJ....132.1221H}
}

@ARTICLE{Dufour2005,
       author = {{Dufour}, P. and {Bergeron}, P. and {Fontaine}, G.},
        title = "{Detailed Spectroscopic and Photometric Analysis of DQ White Dwarfs}",
      journal = {\apj},
         year = 2005,
        month = jul,
       volume = {627},
       number = {1},
        pages = {404-417},
          doi = {10.1086/430373},
archivePrefix = {arXiv},
       eprint = {astro-ph/0503112},
 primaryClass = {astro-ph},
       adsurl = {https://ui.adsabs.harvard.edu/abs/2005ApJ...627..404D}
}

@ARTICLE{Dufour2007,
       author = {{Dufour}, P. and {Bergeron}, P. and {Liebert}, James and {Harris}, H.~C. and {Knapp}, G.~R. and {Anderson}, S.~F. and {Hall}, Patrick B. and {Strauss}, Michael A. and {Collinge}, Matthew J. and {Edwards}, Matt C.},
        title = "{On the Spectral Evolution of Cool, Helium-Atmosphere White Dwarfs: Detailed Spectroscopic and Photometric Analysis of DZ Stars}",
      journal = {\apj},
         year = 2007,
        month = jul,
       volume = {663},
       number = {2},
        pages = {1291-1308},
          doi = {10.1086/518468},
archivePrefix = {arXiv},
       eprint = {astro-ph/0703758},
 primaryClass = {astro-ph},
       adsurl = {https://ui.adsabs.harvard.edu/abs/2007ApJ...663.1291D}
}

@ARTICLE{Bedard2024,
       author = {{B{\'e}dard}, Antoine},
        title = "{The spectral evolution of white dwarfs: where do we stand?}",
      journal = {\apss},
         year = 2024,
        month = apr,
       volume = {369},
       number = {4},
          eid = {43},
        pages = {43},
          doi = {10.1007/s10509-024-04307-5},
archivePrefix = {arXiv},
       eprint = {2405.01268},
 primaryClass = {astro-ph.SR},
       adsurl = {https://ui.adsabs.harvard.edu/abs/2024Ap&SS.369...43B}
}

@ARTICLE{York2000,
       author = {{York}, Donald G. and {Adelman}, J. and {Anderson}, John E., Jr. and {Anderson}, Scott F. and {Annis}, James and {Bahcall}, Neta A. and {Bakken}, J.~A. and {Barkhouser}, Robert and {Bastian}, Steven and {Berman}, Eileen and {Boroski}, William N. and {Bracker}, Steve and {Briegel}, Charlie and {Briggs}, John W. and {Brinkmann}, J. and {Brunner}, Robert and {Burles}, Scott and {Carey}, Larry and {Carr}, Michael A. and {Castander}, Francisco J. and {Chen}, Bing and {Colestock}, Patrick L. and {Connolly}, A.~J. and {Crocker}, J.~H. and {Csabai}, Istv{\'a}n and {Czarapata}, Paul C. and {Davis}, John Eric and {Doi}, Mamoru and {Dombeck}, Tom and {Eisenstein}, Daniel and {Ellman}, Nancy and {Elms}, Brian R. and {Evans}, Michael L. and {Fan}, Xiaohui and {Federwitz}, Glenn R. and {Fiscelli}, Larry and {Friedman}, Scott and {Frieman}, Joshua A. and {Fukugita}, Masataka and {Gillespie}, Bruce and {Gunn}, James E. and {Gurbani}, Vijay K. and {de Haas}, Ernst and {Haldeman}, Merle and {Harris}, Frederick H. and {Hayes}, J. and {Heckman}, Timothy M. and {Hennessy}, G.~S. and {Hindsley}, Robert B. and {Holm}, Scott and {Holmgren}, Donald J. and {Huang}, Chi-hao and {Hull}, Charles and {Husby}, Don and {Ichikawa}, Shin-Ichi and {Ichikawa}, Takashi and {Ivezi{\'c}}, {\v{Z}}eljko and {Kent}, Stephen and {Kim}, Rita S.~J. and {Kinney}, E. and {Klaene}, Mark and {Kleinman}, A.~N. and {Kleinman}, S. and {Knapp}, G.~R. and {Korienek}, John and {Kron}, Richard G. and {Kunszt}, Peter Z. and {Lamb}, D.~Q. and {Lee}, B. and {Leger}, R. French and {Limmongkol}, Siriluk and {Lindenmeyer}, Carl and {Long}, Daniel C. and {Loomis}, Craig and {Loveday}, Jon and {Lucinio}, Rich and {Lupton}, Robert H. and {MacKinnon}, Bryan and {Mannery}, Edward J. and {Mantsch}, P.~M. and {Margon}, Bruce and {McGehee}, Peregrine and {McKay}, Timothy A. and {Meiksin}, Avery and {Merelli}, Aronne and {Monet}, David G. and {Munn}, Jeffrey A. and {Narayanan}, Vijay K. and {Nash}, Thomas and {Neilsen}, Eric and {Neswold}, Rich and {Newberg}, Heidi Jo and {Nichol}, R.~C. and {Nicinski}, Tom and {Nonino}, Mario and {Okada}, Norio and {Okamura}, Sadanori and {Ostriker}, Jeremiah P. and {Owen}, Russell and {Pauls}, A. George and {Peoples}, John and {Peterson}, R.~L. and {Petravick}, Donald and {Pier}, Jeffrey R. and {Pope}, Adrian and {Pordes}, Ruth and {Prosapio}, Angela and {Rechenmacher}, Ron and {Quinn}, Thomas R. and {Richards}, Gordon T. and {Richmond}, Michael W. and {Rivetta}, Claudio H. and {Rockosi}, Constance M. and {Ruthmansdorfer}, Kurt and {Sandford}, Dale and {Schlegel}, David J. and {Schneider}, Donald P. and {Sekiguchi}, Maki and {Sergey}, Gary and {Shimasaku}, Kazuhiro and {Siegmund}, Walter A. and {Smee}, Stephen and {Smith}, J. Allyn and {Snedden}, S. and {Stone}, R. and {Stoughton}, Chris and {Strauss}, Michael A. and {Stubbs}, Christopher and {SubbaRao}, Mark and {Szalay}, Alexander S. and {Szapudi}, Istvan and {Szokoly}, Gyula P. and {Thakar}, Anirudda R. and {Tremonti}, Christy and {Tucker}, Douglas L. and {Uomoto}, Alan and {Vanden Berk}, Dan and {Vogeley}, Michael S. and {Waddell}, Patrick and {Wang}, Shu-i. and {Watanabe}, Masaru and {Weinberg}, David H. and {Yanny}, Brian and {Yasuda}, Naoki and {SDSS Collaboration}},
        title = "{The Sloan Digital Sky Survey: Technical Summary}",
      journal = {\aj},
         year = 2000,
        month = sep,
       volume = {120},
       number = {3},
        pages = {1579-1587},
          doi = {10.1086/301513},
archivePrefix = {arXiv},
       eprint = {astro-ph/0006396},
 primaryClass = {astro-ph},
       adsurl = {https://ui.adsabs.harvard.edu/abs/2000AJ....120.1579Y}
}

@ARTICLE{Chambers2016,
       author = {{Chambers}, K.~C. and {Magnier}, E.~A. and {Metcalfe}, N. and {Flewelling}, H.~A. and {Huber}, M.~E. and {Waters}, C.~Z. and {Denneau}, L. and {Draper}, P.~W. and {Farrow}, D. and {Finkbeiner}, D.~P. and {Holmberg}, C. and {Koppenhoefer}, J. and {Price}, P.~A. and {Rest}, A. and {Saglia}, R.~P. and {Schlafly}, E.~F. and {Smartt}, S.~J. and {Sweeney}, W. and {Wainscoat}, R.~J. and {Burgett}, W.~S. and {Chastel}, S. and {Grav}, T. and {Heasley}, J.~N. and {Hodapp}, K.~W. and {Jedicke}, R. and {Kaiser}, N. and {Kudritzki}, R. -P. and {Luppino}, G.~A. and {Lupton}, R.~H. and {Monet}, D.~G. and {Morgan}, J.~S. and {Onaka}, P.~M. and {Shiao}, B. and {Stubbs}, C.~W. and {Tonry}, J.~L. and {White}, R. and {Ba{\~n}ados}, E. and {Bell}, E.~F. and {Bender}, R. and {Bernard}, E.~J. and {Boegner}, M. and {Boffi}, F. and {Botticella}, M.~T. and {Calamida}, A. and {Casertano}, S. and {Chen}, W. -P. and {Chen}, X. and {Cole}, S. and {Deacon}, N. and {Frenk}, C. and {Fitzsimmons}, A. and {Gezari}, S. and {Gibbs}, V. and {Goessl}, C. and {Goggia}, T. and {Gourgue}, R. and {Goldman}, B. and {Grant}, P. and {Grebel}, E.~K. and {Hambly}, N.~C. and {Hasinger}, G. and {Heavens}, A.~F. and {Heckman}, T.~M. and {Henderson}, R. and {Henning}, T. and {Holman}, M. and {Hopp}, U. and {Ip}, W. -H. and {Isani}, S. and {Jackson}, M. and {Keyes}, C.~D. and {Koekemoer}, A.~M. and {Kotak}, R. and {Le}, D. and {Liska}, D. and {Long}, K.~S. and {Lucey}, J.~R. and {Liu}, M. and {Martin}, N.~F. and {Masci}, G. and {McLean}, B. and {Mindel}, E. and {Misra}, P. and {Morganson}, E. and {Murphy}, D.~N.~A. and {Obaika}, A. and {Narayan}, G. and {Nieto-Santisteban}, M.~A. and {Norberg}, P. and {Peacock}, J.~A. and {Pier}, E.~A. and {Postman}, M. and {Primak}, N. and {Rae}, C. and {Rai}, A. and {Riess}, A. and {Riffeser}, A. and {Rix}, H.~W. and {R{\"o}ser}, S. and {Russel}, R. and {Rutz}, L. and {Schilbach}, E. and {Schultz}, A.~S.~B. and {Scolnic}, D. and {Strolger}, L. and {Szalay}, A. and {Seitz}, S. and {Small}, E. and {Smith}, K.~W. and {Soderblom}, D.~R. and {Taylor}, P. and {Thomson}, R. and {Taylor}, A.~N. and {Thakar}, A.~R. and {Thiel}, J. and {Thilker}, D. and {Unger}, D. and {Urata}, Y. and {Valenti}, J. and {Wagner}, J. and {Walder}, T. and {Walter}, F. and {Watters}, S.~P. and {Werner}, S. and {Wood-Vasey}, W.~M. and {Wyse}, R.},
        title = "{The Pan-STARRS1 Surveys}",
      journal = {arXiv e-prints},
         year = 2016,
        month = dec,
          eid = {arXiv:1612.05560},
        pages = {arXiv:1612.05560},
          doi = {10.48550/arXiv.1612.05560},
archivePrefix = {arXiv},
       eprint = {1612.05560},
 primaryClass = {astro-ph.IM},
       adsurl = {https://ui.adsabs.harvard.edu/abs/2016arXiv161205560C}
}

@ARTICLE{McCleery2020,
       author = {{McCleery}, Jack and {Tremblay}, Pier-Emmanuel and {Gentile Fusillo}, Nicola Pietro and {Hollands}, Mark A. and {G{\"a}nsicke}, Boris T. and {Izquierdo}, Paula and {Toonen}, Silvia and {Cunningham}, Tim and {Rebassa-Mansergas}, Alberto},
        title = "{Gaia white dwarfs within 40 pc II: the volume-limited Northern hemisphere sample}",
      journal = {\mnras},
         year = 2020,
        month = dec,
       volume = {499},
       number = {2},
        pages = {1890-1908},
          doi = {10.1093/mnras/staa2030},
archivePrefix = {arXiv},
       eprint = {2006.00874},
 primaryClass = {astro-ph.SR},
       adsurl = {https://ui.adsabs.harvard.edu/abs/2020MNRAS.499.1890M}
}

@ARTICLE{Caron2023,
       author = {{Caron}, Alexandre and {Bergeron}, P. and {Blouin}, Simon and {Leggett}, S.~K.},
        title = "{A spectrophotometric analysis of cool white dwarfs in the Gaia and pan-STARRS footprint}",
      journal = {\mnras},
         year = 2023,
        month = mar,
       volume = {519},
       number = {3},
        pages = {4529-4549},
          doi = {10.1093/mnras/stac3733},
archivePrefix = {arXiv},
       eprint = {2212.08014},
 primaryClass = {astro-ph.SR},
       adsurl = {https://ui.adsabs.harvard.edu/abs/2023MNRAS.519.4529C}
}

@ARTICLE{OBrien2024,
       author = {{O'Brien}, Mairi W. and {Tremblay}, P. -E. and {Klein}, B.~L. and {Koester}, D. and {Melis}, C. and {B{\'e}dard}, A. and {Cukanovaite}, E. and {Cunningham}, T. and {Doyle}, A.~E. and {G{\"a}nsicke}, B.~T. and {Gentile Fusillo}, N.~P. and {Hollands}, M.~A. and {McCleery}, J. and {Pelisoli}, I. and {Toonen}, S. and {Weinberger}, A.~J. and {Zuckerman}, B.},
        title = "{The 40 pc sample of white dwarfs from Gaia}",
      journal = {\mnras},
         year = 2024,
        month = jan,
       volume = {527},
       number = {3},
        pages = {8687-8705},
          doi = {10.1093/mnras/stad3773},
archivePrefix = {arXiv},
       eprint = {2312.02735},
 primaryClass = {astro-ph.SR},
       adsurl = {https://ui.adsabs.harvard.edu/abs/2024MNRAS.527.8687O}
}

@ARTICLE{vcs73,
       author = {{Vidal}, C.~R. and {Cooper}, J. and {Smith}, E.~W.},
        title = "{Hydrogen Stark-Broadening Tables}",
      journal = {\apjs},
         year = 1973,
        month = jan,
       volume = {25},
        pages = {37},
          doi = {10.1086/190264},
       adsurl = {https://ui.adsabs.harvard.edu/abs/1973ApJS...25...37V}
}

@BOOK{mihalas78,
       author = {{Mihalas}, Dimitri},
        title = "{Stellar Atmospheres}",
         year = 1978,
    publisher = {San Francisco: W. H. Freeman},
       adsurl = {https://ui.adsabs.harvard.edu/abs/1978stat.book.....M}
}

@ARTICLE{Cho2022,
       author = {{Cho}, P.~B. and {Gomez}, T.~A. and {Montgomery}, M.~H. and {Dunlap}, B.~H. and {Fitz Axen}, M. and {Hobbs}, B. and {Hubeny}, I. and {Winget}, D.~E.},
        title = "{Simulation of Stark-broadened Hydrogen Balmer-line Shapes for DA White Dwarf Synthetic Spectra}",
      journal = {\apj},
         year = 2022,
        month = mar,
       volume = {927},
       number = {1},
          eid = {70},
        pages = {70},
          doi = {10.3847/1538-4357/ac4df3},
       adsurl = {https://ui.adsabs.harvard.edu/abs/2022ApJ...927...70C}
}

@ARTICLE{Rosato2020,
       author = {{Rosato}, J. and {Marandet}, Y. and {Stamm}, R.},
        title = "{Quantifying the statistical noise in computer simulations of Stark broadening}",
      journal = {\jqsrt},
         year = 2020,
        month = jul,
       volume = {249},
          eid = {107002},
        pages = {107002},
          doi = {10.1016/j.jqsrt.2020.107002},
       adsurl = {https://ui.adsabs.harvard.edu/abs/2020JQSRT.24907002R}
}

@ARTICLE{Hooper68,
       author = {{Hooper}, C.~F.},
        title = "{Asymptotic Electric Microfield Distributions in Low-Frequency Component Plasmas}",
      journal = {Physical Review},
         year = "1968",
        month = "May",
       volume = {169},
       number = {1},
        pages = {193-195},
          doi = {10.1103/PhysRev.169.193},
       adsurl = {https://ui.adsabs.harvard.edu/abs/1968PhRv..169..193H}
}

@ARTICLE{Kilic2025,
       author = {{Kilic}, Mukremin and {Bergeron}, Pierre and {Blouin}, Simon and {Moss}, Adam and {Brown}, Warren R. and {B{\'e}dard}, Antoine and {Jewett}, Gracyn and {Ag{\"u}eros}, Marcel A.},
        title = "{The 100 pc White Dwarf Sample in the SDSS Footprint. II. A New Look at the Spectral Evolution of White Dwarfs}",
      journal = {\apj},
         year = 2025,
        month = feb,
       volume = {979},
       number = {2},
          eid = {157},
        pages = {157},
          doi = {10.3847/1538-4357/ad9bb3},
       adsurl = {https://ui.adsabs.harvard.edu/abs/2025ApJ...979..157K}
}

@ARTICLE{Lara12,
       author = {{Lara}, N. and {Gonz{\'a}lez}, M. {\'A}. and {Gigosos}, M.~A.},
        title = "{Stark broadening tables for the helium I 492.2 line. Application to weakly coupled plasma diagnostics}",
      journal = {\aap},
         year = 2012,
        month = jun,
       volume = {542},
          eid = {A75},
        pages = {A75},
          doi = {10.1051/0004-6361/201219123},
       adsurl = {https://ui.adsabs.harvard.edu/abs/2012A&A...542A..75L}
}

@INPROCEEDINGS{Gomez2017,
       author = {{Gomez}, T.~A. and {Montgomery}, M.~H. and {Nagayama}, T. and {Kilcrease}, D.~P. and {Winget}, D.~E.},
        title = "{Modeling the Spectra of Dense Hydrogen Plasmas: Beyond Occupation Probability}",
    booktitle = {20th European White Dwarf Workshop},
         year = 2017,
       editor = {{Tremblay}, P. -E. and {Gaensicke}, B. and {Marsh}, T.},
       series = {Astronomical Society of the Pacific Conference Series},
       volume = {509},
        month = mar,
        pages = {143},
          doi = {10.48550/arXiv.1610.02342},
archivePrefix = {arXiv},
       eprint = {1610.02342},
 primaryClass = {astro-ph.SR},
       adsurl = {https://ui.adsabs.harvard.edu/abs/2017ASPC..509..143G}
}

@ARTICLE{Baranger62,
       author = {{Baranger}, M.},
        title = "{Spectral Line Broadening in Plasmas}",
      journal = {Pure and Applied Physics},
         year = "1962",
        month = "Jan",
       volume = {13},
        pages = {493-548},
          doi = {10.1016/B978-0-12-081450-3.50017-5},
       adsurl = {https://ui.adsabs.harvard.edu/abs/1962PApPh..13..493B}
}

@ARTICLE{Tremblay2026,
       author = {{Tremblay}, Patrick and {Beauchamp}, Alain and {Bergeron}, Pierre and {B{\'e}dard}, Antoine},
        title = "{Improved Stark-broadened Profiles for Neutral Helium Lines Using Computer Simulations}",
      journal = {\apj},
         year = 2026,
        month = apr,
       volume = {1000},
       number = {2},
          eid = {253},
        pages = {253},
          doi = {10.3847/1538-4357/ae4e26},
archivePrefix = {arXiv},
       eprint = {2603.04374},
 primaryClass = {astro-ph.SR},
       adsurl = {https://ui.adsabs.harvard.edu/abs/2026ApJ..1000..253T}
}

@ARTICLE{Dappen1987,
       author = {{D{\"a}ppen}, Werner and {Anderson}, Lawrence and {Mihalas}, Dimitri},
        title = "{Statistical Mechanics of Partially Ionized Stellar Plasmas: The Planck-Larkin Partition Function, Polarization Shifts, and Simulations of Optical Spectra}",
      journal = {\apj},
         year = 1987,
        month = aug,
       volume = {319},
        pages = {195},
          doi = {10.1086/165446},
       adsurl = {https://ui.adsabs.harvard.edu/abs/1987ApJ...319..195D}
}

@ARTICLE{Bergeron1991,
       author = {{Bergeron}, P. and {Wesemael}, F. and {Fontaine}, G.},
        title = "{Synthetic Spectra and Atmospheric Properties of Cool DA White Dwarfs}",
      journal = {\apj},
         year = 1991,
        month = jan,
       volume = {367},
        pages = {253},
          doi = {10.1086/169624},
       adsurl = {https://ui.adsabs.harvard.edu/abs/1991ApJ...367..253B}
}

\end{document}